%% file: main.tex
\documentclass[aps,prd,nofootinbib,aps,tightenlines,preprintnumbers,notitlepage,longbibliography,superscriptaddress]{revtex4-2}
\usepackage{graphicx}
\usepackage{amsmath}
\usepackage{amsfonts}
\usepackage[colorlinks=true,citecolor=blue,urlcolor=black,filecolor=black,runcolor=black,linkcolor=black]{hyperref}
\usepackage{appendix}
\usepackage{setspace}
\usepackage{slashed}
\usepackage{bm}
\usepackage{titlesec}

\titlespacing*{\section}{0pt}{1.1\baselineskip}{0.6\baselineskip}
\titlespacing*{\subsection}{0pt}{0.9\baselineskip}{0.45\baselineskip}
\titlespacing*{\subsubsection}{0pt}{0.75\baselineskip}{0.35\baselineskip}

\usepackage{tabularx}
\newcolumntype{C}{>{\centering\arraybackslash}X}
\newcolumntype{R}{>{\raggedleft\arraybackslash}X}

\DeclareRobustCommand{\fnl}{f_{\rm NL}}

\newcommand{\Pgg}{P_{gg}}
\newcommand{\Pgv}{P_{gv}}
\newcommand{\Pvv}{P_{vv}}
\newcommand{\hPgg}{\hat{P}_{gg}}

\newcommand{\bvel}{b_v}
\newcommand{\bfg}{b_{\rm fg}}
\newcommand{\sigmav}{\Sigma_v}

\usepackage[usenames, dvipsnames]{color}
\usepackage[normalem]{ulem}
\usepackage{xcolor}

\definecolor{vchR}{HTML}{C2185B}
\definecolor{vchG}{HTML}{008A2E}
\definecolor{vchP}{HTML}{7B1FA2}
\definecolor{vchC}{HTML}{0097A7}
\definecolor{vchS}{HTML}{E65100}

\AtBeginDocument{\pdfstringdefDisableCommands{}}
\usepackage{fontawesome5}
\usepackage{orcidlink}

\def\bx{{\boldsymbol{x}}}
\def\bell{{\boldsymbol{\ell}}}
\def\btheta{{\boldsymbol{\theta}}}

\def\bd{{\boldsymbol{d}}}

\newcommand{\bk}{\boldsymbol{k}}

\newcommand{\be}{\begin{eqnarray}}

\newcommand{\ee}{\end{eqnarray}}

\definecolor{firebrick}{HTML}{B22222}
\definecolor{darkgreen}{HTML}{006400}

\usepackage{multirow}
\usepackage{booktabs}

\usepackage{dblfloatfix}

\AtBeginDocument{\raggedbottom}

\definecolor{sfdone}{HTML}{0097A7}

\newcommand{\perimeter}{Perimeter Institute for Theoretical Physics, 31 Caroline St N, Waterloo, ON N2L 2Y5, Canada}

\graphicspath{ {plots/} }

\begin{document}

\title{Velocity Reconstruction from kSZ with SPT-3G and DESI-LS}

\author{Selim~C.~Hotinli\orcidlink{0000-0003-0061-8188}}
\email{shotinli@perimeterinstitute.ca}
\affiliation{\perimeter}

\author{Edmond~Chaussidon\orcidlink{0000-0001-8996-4874}}
\affiliation{Lawrence Berkeley National Laboratory, 1 Cyclotron Road, Berkeley, CA 94720, USA}

\author{Simone~Ferraro\orcidlink{0000-0003-4992-7854}}
\affiliation{Lawrence Berkeley National Laboratory, 1 Cyclotron Road, Berkeley, CA 94720, USA}
\affiliation{Berkeley Center for Cosmological Physics, Department of Physics, University of California, Berkeley, CA 94720, USA}

\author{Kendrick~M.~Smith\orcidlink{0000-0002-2088-3125}}
\affiliation{\perimeter}

\begin{abstract}

We reconstruct the large-scale radial velocity field from the kinetic Sunyaev--Zel'dovich effect for the first time with the South Pole Telescope, and measure the galaxy--velocity correlation and the velocity power spectrum. We cross-correlate the SPT-3G D1 maps at 95, 150, and 220\,GHz with luminous red galaxies from the DESI Legacy Imaging Surveys over the SPT-3G Main field. For the extended galaxy sample, we detect the galaxy--velocity correlation at signal-to-noise ratio $11.2$ and the velocity power spectrum at $8.1$, combining the three frequencies. We detect the reconstructed velocities for the first time at 220\,GHz with signal-to-noise ratios of 4.2 and 3.1 from the galaxy--velocity correlation dipole and monopole, and at 3.8 from the velocity power spectrum. {We model and estimate foreground contributions to the monopole, detecting both the thermal Sunyaev Zel'dovich effect and the cosmic infrared background contribution separately. On the around five percent of sky jointly covered by SPT-3G and DESI Legacy Imaging Surveys,} reconstructed velocities give new constraints on local-type primordial non-Gaussianity at $\fnl=-38^{+61}_{-61}$, and adding galaxy clustering gives $\fnl=-20^{+40}_{-46}$. The galaxy--velocity correlation is consistent between the three frequencies. \end{abstract}

\maketitle

\section{Introduction}\label{sec:intro}

The large-scale peculiar-velocity field probes the matter fluctuations and traces the growth of cosmic structure, providing a direct test of cosmological models~\citep[see~e.g.][]{Strauss:1995fz,Bertschinger:1993xt,Koda:2013eya}. The kinetic Sunyaev-Zel'dovich (kSZ) effect is the Doppler shift that cosmic microwave background (CMB) photons acquire when they scatter off free electrons with a bulk peculiar velocity~\cite{Sunyaev:1980nv,Ostriker:1986xc}. The kSZ effect, therefore, probes the line-of-sight component of this velocity field. In recent years, velocity reconstruction from the kSZ effect has become a competitive probe of large-scale structure~\cite{Smith:2018bpn, Munchmeyer2018, Krywonos:2024mpb, Lague:2024czc, McCarthy:2024nik, Hotinli:2025tul, Lai:2025qdw, McCarthy:2025ksz, Gandhi2026, EmbilVillagra2026, Chaussidon:2026vmn}.

The kSZ effect has been measured through several complementary techniques. Pairwise estimators stack cluster pairs to detect their mean relative radial motion, first in Atacama Cosmology Telescope (ACT) and Sloan Digital Sky Survey (SDSS)~\cite{Hand2012}, and subsequently in South Pole Telescope (SPT) and Dark Energy Survey (DES)~\cite{DES:2016umt}, ACT~\cite{Calafut:2021wkx}, SPT-3G and DES~\cite{SPT-3G:2022zrq}, and ACT, \textit{Planck} and Dark Energy Spectroscopic Instrument (DESI)~\cite{Gong:2025ffw}. Projected-field estimators cross-correlate the squared CMB temperature with a projected galaxy field~\cite{Hill:2016dta, Ferraro:2016ymw, Kusiak:2021hai}.
The small-scale temperature power spectrum constrains the total kSZ power~\cite{SPT:2020psp, ACT:2020frw, SPT-3G:2026mcr}, and the kSZ trispectrum~\cite{Smith:2016lnt} sets an upper limit on the duration of reionization~\cite{SPT-3G:2024lko}.
Velocity-weighted stacks of CMB maps on galaxies trace the gas density profiles around them~\cite{ACTPol:2015teu, AtacamaCosmologyTelescope:2020wtv, Hadzhiyska:2024qsl, RiedGuachalla:2025byu}, and Fourier-space estimators have measured the galaxy--electron cross-power spectrum $P_{ge}$~\cite{Qu:2026zyh, Hadzhiyska:2026uyq, Hotinli:2026zne}.

This work uses kSZ \emph{velocity reconstruction}, which recovers the three-dimensional radial peculiar-velocity field on large scales from the correlation between the small-scale CMB temperature and the small-scale galaxy density~\cite{Deutsch:2017ybc,Smith:2018bpn,Munchmeyer2018,Cayuso:2021ljq,Kvasiuk:2023nje}. We cross-correlate the reconstructed velocity field with the galaxy density field, measuring the bispectrum of two galaxy modes and the CMB temperature~(\citet{Smith:2018bpn}), which $P_{ge}$ estimators probe on small scales~\cite[see e.g.][]{Hotinli:2026zne} and velocity reconstruction on large scales.

Beyond detecting the kSZ signal, velocity reconstruction also helps us probe the initial conditions of the Universe through primordial non-Gaussianity (PNG). Local form of PNG induces a scale-dependent bias in the galaxy density field~\citep{Dalal:2007cu}. The galaxy-velocity cross-power $\Pgv$ is sensitive to $\fnl$ through this bias, which is largest on the largest scales. On these scales the reconstructed velocity is a second tracer of the matter density and, through sample-variance cancellation~\cite{Seljak:2008xr}, can sharpen the galaxy-clustering constraint on $\fnl$~\cite{Dalal:2007cu,Munchmeyer2018}. The tightest $\fnl$ constraints come from CMB bispectra (e.g.\ \textit{Planck}'s $\fnl=-0.9\pm5.1$~\cite{Planck:2019kim}) and the large-scale structure offers a complementary, low-redshift approach: the strongest bounds come from the scale-dependent clustering of spectroscopic DESI samples, with $\fnl = -3.6^{+9.0}_{-9.1}$ from the combined DESI~DR1 luminous-red-galaxy (LRG) and quasar power spectra~\cite{Chaussidon:2024qni}. The current bounds on $\fnl$ from velocity reconstruction are at $\pm34$ (\citet{Chaussidon:2024qni}), where this gap is expected to be closed in the near future with Simons Observatory. 

The kSZ velocity reconstruction program has experienced a recent boom in the past two years. \citet{Bloch:2024kpn}~and~\citet{Krywonos:2024mpb} performed a light-cone reconstruction from \textit{Planck} and unWISE data, placed an upper limit on the amplitude of the reconstructed signal, and constrained $-220\lesssim\fnl\lesssim136$ at $68\%$ confidence. \citet{McCarthy:2024nik} first achieved a statistically significant velocity reconstruction from the ACT~DR6 CMB maps and photometric DESI LRGs, detecting the signal at a signal-to-noise ratio $\mathrm{SNR} = 3.8$ by cross-correlating with an external estimate of the velocity field from the Baryon Oscillation Spectroscopic Survey (BOSS).
\citet{Lague:2024czc} measured $\Pgv$ at $\mathrm{SNR} = 7.2$ with ACT~DR5 and spectroscopic SDSS. \citet{Hotinli:2025tul} presented the first three-dimensional reconstruction of the large-scale velocity field using a photometric galaxy survey, from ACT~DR5 and DESI Legacy Imaging Surveys (DESI-LS) LRGs reaching detection $\mathrm{SNR} = 11.7$ with the extended sample and the day$+$night maps. \citet{Lai:2025qdw} detected the signal at $\mathrm{SNR}\sim10$ with ACT~DR6 $\times$ DESI-LS~DR9 and a quadratic-maximum-likelihood power spectrum estimator. \citet{McCarthy:2025ksz} cross-correlated a kSZ velocity reconstruction, from ACT~DR6 and photometric DESI LRGs, with the velocity inferred from the continuity equation applied to the same galaxies, at $\mathrm{SNR} \sim 10$ ($\sim 11$ combined with the velocity auto-correlation). \citet{EmbilVillagra2026} modelled the thermal Sunyaev--Zel'dovich (tSZ) and cosmic infrared background (CIB) biases to the projected galaxy--velocity cross-correlation and introduced a parity-odd estimator, built from antisymmetric combinations of tomographic galaxy--velocity correlations, that cancels these foregrounds in the Limber approximation while preserving the signal. Most recently, \citet{Chaussidon:2026vmn} combined DESI~DR2 with ACT~DR6, reporting a total $\mathrm{SNR}$ of $20.8$ across LRG, emission-line galaxy and quasar samples (17.0, 8.3 and 6.8 per tracer), which is the highest significance $\Pgv$ detection to date, and the first detection of the velocity auto-spectrum $\Pvv$, at $\mathrm{SNR} = 3.1$.

In this work, we use the SPT-3G D1 maps and DESI-LS to measure the kSZ effect. Previously SPT and DES data have been combined in pairwise kSZ~\cite{DES:2016umt, SPT-3G:2022zrq} measurements, in cross correlations of CMB lensing and galaxy clustering~\cite{DES:2018miq, DES:2018fvb, DES:2022qdz, DES:2022urg}, cluster cosmology with weak-lensing mass calibration~\cite{DES:2023muu,DES:2024viv}, and CMB-cluster lensing~\cite{SPT:2019qkp,SPT-3G:2024lri} measurements; but not yet in kSZ velocity reconstruction. \citet{Hotinli:2026zne} measured $P_{ge}$ with the SPT-3G D1 maps and DESI-LS LRGs, weighting the galaxies by radial velocities reconstructed from the galaxy density rather than from the CMB.

We measure $\Pgv$ and $\Pvv$ with SPT-3G and DESI-LS, using the SPT-3G D1 maps at 95, 150, and 220\,GHz~\cite{SPT-3G:2026krg} and the main and extended DESI-LS LRG samples of Ref.~\citep{Zhou:2023gji}.
Using the main sample, we detect the $\Pgv^{\ell=1}$ dipole at SNR $10.4$ over the three frequencies, the velocity auto-spectrum at $7.2$. The SNR values for the extended sample are 11.2 and 8.1, respectively. We measure the foreground-induced monopole at $16.8$ (18.9) for the main (extended) sample.
The velocity spectra give $\fnl=-38^{+61}_{-61}$, and adding the galaxy auto-spectrum gives $\fnl=-20^{+40}_{-46}$, both consistent with zero, as the previous ACT and DESI measurements~\citep{Hotinli:2025tul, Chaussidon:2026vmn} that used a similar pipeline.
The frequency differences of the dipole are consistent with zero, and the extended sample gives results that agree with the main sample.

The remainder of this paper is organized as follows. In \S\ref{sec:pipeline} we describe the pipeline that produces the $\Pgv$ dipole and monopole and the velocity auto-spectrum $\Pvv$. In \S\ref{sec:data}, we describe our data: the SPT-3G D1 maps and the DESI-LS LRG catalogs. In \S\ref{sec:results}, we present the results, including the galaxy auto-spectrum, the galaxy-velocity correlations, and velocity spectra, $\fnl$, as well as the consistency tests; and in \S\ref{sec:discussion}, we conclude. Our pipeline, methods, and analysis follow the Refs.~\cite{Hotinli:2025tul, Chaussidon:2026vmn} closely.

\section{Preliminaries}\label{sec:prelim}

Throughout this paper, $\bx$ is a three-dimensional comoving position, the unit three-vector $\btheta$ is a direction on the sky, and $\hat\bx$ is the unit three-vector along the line of sight, pointing away from the observer. We write the comoving distance to redshift $z$ as $\chi(z)$ and switch between $\chi$ and $z$ as the radial variable where convenient. We use the following Fourier convention:
\begin{equation}\label{eq:fourier_conv}
f(\bx) = \int \frac{d^3\bk}{(2\pi)^3}\, \tilde f(\bk)\, e^{i\bk\cdot\bx}\,, \hspace{1.5cm} \tilde f(\bk) = \int d^3\bx\, f(\bx)\, e^{-i\bk\cdot\bx}\,,
\end{equation}
where $\bk$ is the three-dimensional Fourier wavevector, and we define the wavenumber as $k=|\bk|$ in what follows.

We write the kSZ contribution to the CMB temperature as a line-of-sight integral~\citep{Sunyaev:1980nv}:
\begin{equation}\label{eq:tksz_line_of_sight}
T_{\rm kSZ}(\btheta) = \int d\chi\, K(\chi)\, v_r(\chi\btheta)\, \delta_e(\chi\btheta)\,,
\end{equation}
with $\delta_e(\chi\btheta)$ and $v_r(\chi\btheta)$ the electron overdensity and the radial velocity at comoving distance $\chi$ in the direction $\btheta$. The radial weight $K(z)$ is given by:
\begin{equation}\label{eq:Kz}
K(z) \equiv -T_{\rm CMB}\, \sigma_T\, n_{e,0}\, x_e(z)\, e^{-\tau(z)}\, (1+z)^2\,,
\end{equation}
where $T_{\rm CMB}$ and $n_{e,0}$ are the present-day mean CMB temperature and mean electron density, $\sigma_T$ is the Thomson cross-section, and $x_e(z)$ and $\tau(z)$ are the ionization fraction and the optical depth to redshift $z$. Throughout this work, we take a fixed redshift at the mean of our galaxy sample $z_*=0.754$, and set $x_e=1$ and $e^{-\tau}=1$.

In the presence of primordial non-Gaussianity (PNG), the galaxy bias $b_g$ gains an extra term on large scales~\citep{Dalal:2007cu, Slosar:2008hx}:
\begin{equation}\label{eq:scale_dep_bias}
b(k) = b_g + 2\delta_c(b_g - 1)\,\frac{\fnl}{\alpha(k,z)}\,, \hspace{1.5cm} \alpha(k,z) = \frac{\delta_m(\bk,z)}{\Phi(\bk,z)} = \frac{2 k^2 T(k) D(z)}{3\,\Omega_{m0}\, H_0^2}\,,
\end{equation}
where $T(k)$ and $D(z)$ are the matter transfer function and the linear growth factor, $\Omega_{m0}$ and $H_0$ the present-day matter density and the Hubble constant, and we adopt the spherical-collapse threshold $\delta_c=1.68$. We evaluate $\alpha(k,z)$ directly from the linear matter power spectrum.
Since $\alpha(k,z)\propto k^2$ as $k\rightarrow0$, the PNG correction is largest on the largest scales.
The coefficient $2\delta_c(b_g-1)$ assumes a universal mass function and is taken to be equal to the so-called `local PNG bias' $b_\Phi$. Departures from this relation can shift the inferred $\fnl$~\cite{Barreira:2020abc, Barreira:2020bbc}. Recent works has shown,
however, that $b_\Phi$ can be measured from how the number of galaxies changes with redshift~\citep{Dalal:2025eve,Sullivan:2025fie}. In what follows we will assume $b_{\Phi}$ fixed, and measure the $f_{\rm NL}$ from galaxy and galaxy-velocity correlations.

\section{Pipeline}\label{sec:pipeline}

We begin in this section with a description of our velocity-reconstruction framework. The majority of our analysis pipeline was first introduced in \citet{Hotinli:2025tul}, followed by improvements made in \citet{Chaussidon:2026vmn}.

\subsection{kSZ velocity reconstruction estimator}\label{sec:vel_estimator}

The kSZ quadratic estimator reconstructs the large-scale radial-velocity field (up to a multiplicative bias) from a CMB temperature map and a galaxy catalogue. The output of this reconstruction is a 3-d velocity field $\hat v_r(\bx)$ on large scales, whereas the inputs are the 3-d galaxy field $\delta_g(\bx)$ and the 2-d CMB $T(\btheta)$ on small scales. 

The small-scale galaxy field is memory and compute intensive at high-resolution (at wavenumbers $k_S\sim1~{\rm Mpc}^{-1}$), so we define our estimator as a weighted catalog:
\begin{equation}\label{eq:est_definition}
    \hat v_r(\bx)=\sum_{i\in\rm gal}W_i^v\,\tilde T(\btheta_i)\delta^3(\bx-\bx_i)\,,
\end{equation}
where $\btheta_i$ is the angular location of the $i$-th galaxy, $W_i^v$ is a survey-specific per-galaxy weight, and $\tilde T(\btheta)$ is a high-pass filtered CMB map which we construct in harmonic space as:
\begin{equation}\label{eq:filter_temperature}
\tilde T_{\ell m} = F_\ell^\nu \int d^2\btheta\, W_{\rm CMB}(\btheta)\, T^\nu(\btheta)\, Y^*_{\ell m}(\btheta)\,,
\end{equation}
where $T^\nu$ is the temperature map at frequency $\nu$ and $W_{\rm CMB}$ is the CMB pixel weight. Here, both $W_i^v$ and $\tilde{T}(\btheta_i)$ are weights that act on our galaxy catalogs. $F_\ell^v$ is the $\ell$-weighting of the quadratic estimator.

For our purposes, the optimal choice for $F_\ell^v$ down-weights CMB multipoles by the variance of the maps we input; the total galaxy power spectrum (including Poisson noise) $P_{gg}^{\rm tot}$ and the total CMB power spectrum (including foregrounds and noise) $C_\ell^{\rm tot,\nu}$; and up-weights them by galaxy-electron correlation $P_{ge}$. Here, we define the $\ell$-weighting as:
\begin{equation}\label{eq:fl_def}
F_\ell^\nu = \frac{P_{ge}^{\rm fid}(\ell/\chi_*)}{P_{gg}^{\rm fid}(\ell/\chi_*)} \, \frac{b_\ell^\nu}{C_\ell^{{\rm tot},\,\nu}}\,,
\end{equation}
where $P_{ge}^{\rm fid}$ and $P_{gg}^{\rm fid}$ are our fiducial choices for the galaxy power spectrum and galaxy-electron correlation, which we calculate from the halo-model at $k = \ell/\chi_*$, with $\chi_* = \chi(z_*)$,  $b_\ell^v$ is the CMB survey beam, and total observed CMB temperature power spectrum $C_\ell^{{\rm tot},\,\nu}$ is fitted from the data.

The CMB contains other signals, such as those from the thermal Sunyaev-Zel'dovich (tSZ) effect~\citep{Sunyaev1972} and the cosmic infrared background (CIB)~\citep{Hauser2001}, that are correlated with galaxies. These are \textit{foregrounds} for our analysis and contribute to $\tilde T(\btheta_i)$. For the remainder of this section, we omit these contributions and return to the description of our modelling and mitigation of these in~\S\ref{sec:foregrounds}.

The reconstructed velocity in Eq.~\eqref{eq:est_definition} is biased relative to the true radial-velocity field \cite{Smith:2018bpn}. The mean of the reconstructed velocity satisfies~(see~\citet{Hotinli:2025tul} for the derivation):
\begin{equation}\label{eq:hvr_bias}
\langle \hat v_r(\bx) \rangle = \bvel \, B(\bx) \, n_v(\bx) \, v_r^{\rm true}(\bx)\,,
\end{equation}
where $n_v(\bx) = \langle\sum_i W^v_i\,\delta^3(\bx - \bx_i)\rangle$ is the weighted 3-d galaxy number density, $b_v$ is the `kSZ velocity bias', well approximated as scale-independent on large scales $(k\lesssim0.04)$ that we consider in this work, as shown in $N$-body simulations in~Ref.~\citep{Giri:2020pkk}. The function $B(\bx)$ is defined by:
\begin{equation}\label{eq:B_def}
 B(\bx)\equiv W_{\rm CMB}(\btheta) B_0(\chi)\,,\hspace{1.5cm}\textrm{where}\hspace{1.5cm} B_0(\chi)\equiv\frac{K(\chi)}{\chi^2}\, \int \frac{d^2\bell}{(2\pi)^2}\, b_\ell^\nu\, F_\ell^\nu\, P_{ge}^{\rm fid}(\ell/\chi)\,.
\end{equation}
The kSZ velocity bias is formally defined by:
\begin{equation}\label{eq:bv_def}
\bvel(\chi) = \frac{\int (d^2\bell/(2\pi)^2)\, b_\ell^\nu\, F_\ell^\nu\, P_{ge}^{\rm true}(\ell/\chi)}{\int (d^2\bell/(2\pi)^2)\, b_\ell^\nu\, F_\ell^\nu\, P_{ge}^{\rm fid}(\ell/\chi)}\,,
\end{equation}
which is an overall prefactor that is roughly the ratio between the true galaxy-electron power spectrum $P_{ge}^{\rm true}$ and the fiducial galaxy-electron power spectrum we defined above. Both $B_0(\chi)$ and $\bvel(\chi)$ vary slowly with redshift, and following~\citet{Hotinli:2025tul} we fit $\bvel$ as a constant in redshift.

\subsection{Power spectrum estimators}\label{sec:surrogates_model}
\label{sec:pk_est}

Once reconstructing radial velocities following \S\ref{sec:vel_estimator}, we estimate the power spectra of the radial velocity and its cross-correlation with galaxies. We define multi-polar spin-$\ell$ power spectra as in the galaxy clustering analyses. In our case, our main observable—the reconstructed radial velocity field—is a spin-1 field. In this section, we describe how we define its spectra, summarizing the formalism developed in Refs.~\citep{Hotinli:2025tul,Chaussidon:2026vmn}, which builds on the dipolar power-spectrum estimators of Refs.~\cite{Hand:2017irw, Scoccimarro:2015bla}.

For a field $f$, the spin-$\ell$ Fourier transform pair is
\begin{equation}\label{eq:spin_l_ft}
\begin{cases}
\mathcal{F}_\ell[f](\bk) = \displaystyle\int d^3\bx\, \epsilon^*\, \mathcal{L}_\ell(\mu)\, f(\bx)\, e^{-i\bk\cdot\bx}\,, \\[1ex]
\mathcal{F}_\ell[f](\bx) = \displaystyle\int \frac{d^3\bk}{(2\pi)^3}\, \epsilon\, \mathcal{L}_\ell(\mu)\, f(\bk)\, e^{i\bk\cdot\bx}\,,
\end{cases}
\end{equation}
where $\mu = \hat\bk\cdot\hat\bx$, $\mathcal{L}_\ell$ is the Legendre polynomial of order $\ell$, and $\epsilon = i$ if $\ell$ is odd and $\epsilon = 1$ if $\ell$ is even. We use Eq.~\eqref{eq:spin_l_ft} to define
\begin{equation}
    \hat{v}_1 \equiv \mathcal{F}_1[\hat v_r](\bk)=\int d^3\bx\,i\mu\,\hat v_r(\bx)e^{-i\bk\cdot\bx}\hspace{1.5cm}\textrm{and}\hspace{1.5cm}\hat{P}^{\ell=1}_{gv}(k)=\frac{3}{\mathcal{N}_{gv}}\int\frac{d^3\bk}{(2\pi)^3}\rho_g^*(\bk)\hat v_1(\bx)\,,
\end{equation}
where $\mathcal{N}_{gv}$ is a normalization factor~(see Appendix~A~of~\citet{Hotinli:2025tul} for our definitions of power spectra normalization we use throughout this paper) and $\rho_g(\bx)$ is the 3-d large-scale galaxy field, which we define through the `galaxies-minus-randoms' prescription:
\begin{equation}
    \rho_g(\bx)\equiv\sum_{i\in{\rm gal}}W_i^g\delta^3(\bx-\bx_i)-\frac{N_g}{N_r}\sum_{j\in{\rm rand}}W_j^g\delta^3(\bx-\bx_j)\,,
\end{equation}
where $W_i^g$ is a per-galaxy weight (such as the FKP or imaging weights), and $N_g$ and $N_r$ are the number of galaxies in the  data catalog and randoms, respectively. 

The estimator $\hat P_{gv}^{\ell=1}$ of the galaxy-velocity correlation \textit{dipole} is our main observable in this work, which we will compare to a few-parameter model for $P_{gv}(k)$:
\begin{equation}
    P_{gv}(k)=b_v\left(b_g+f_{\rm NL}\frac{2\delta_c(b_g-1)}{\alpha(k,z)}\right)\frac{faH}{k}P_{\rm lin}(k)
\end{equation}
where $P_{\rm lin}(k)$ is the linear matter power spectrum, $f$ is the linear growth rate, $a$ is the scale factor and $H$ is the Hubble expansion rate.

In addition to the galaxy-velocity correlation, we also define power-spectrum estimators for velocity-velocity auto-correlation, 
\begin{equation}\label{eq:Pgv_dipole}
\hat{P}_{vv}(k)=\frac{9}{\mathcal{N}_{vv}}\int\frac{d^3\bk}{(2\pi)^3}F_1[\hat v_r](\bk)F_1[\hat v_r](-\bk)\,,
\end{equation}
and for reasons that will become clear in \S\ref{sec:foregrounds}, the galaxy-velocity cross-correlation \textit{monopole} by:
\begin{equation}\label{eq:Pgv_monopole}
\hat{P}_{gv}^{\ell=0}(k)=\frac{1}{\mathcal{N}_{gv}}\int\frac{d^3\bk}{(2\pi)^3}F_0[\rho_g](\bk)F_0[\hat v_r](-\bk)\,.
\end{equation}
We define the galaxy power spectrum as:
\begin{equation}
    \hat P_{gg}(k)\equiv\frac{1}{\mathcal N_{gg}}\int\frac{d^3\bk}{(2\pi)^3}\rho^*_g(\bk)\rho_g(\bk)\,.
\end{equation}

A standard method in cosmology for comparing these spectra to their theoretical predictions is to compute and de-convolve the effect of survey geometry. In this paper, we use a different approach: we convolve the model for these spectra with the survey geometry and compute the response of the estimators to the model in a way that accounts for the survey geometry.

\subsection{Foreground contributions}\label{sec:foregrounds}

So far in constructing the estimator of \S\ref{sec:vel_estimator} we treated the filtered CMB temperature $\tilde T$ of Eq.\ (\ref{eq:filter_temperature}) as containing only the kSZ signal. In practice, other signals (foregrounds, for our purposes) in the CMB that are correlated with the galaxy survey will also contribute to our estimator. Here we describe how our pipeline accounts for these contributions.

First, we have found in \citet{Hotinli:2025tul} that we have to modify our estimator in order to mitigate correlated CMB foregrounds that on average make an additive contribution to the value of $\tilde T$ in a way that is slowly varying in redshift and independent of the radial velocity. In order to mitigate this systematic contribution, we divide our galaxies into redshift bins $b$ and subtract the mean weighted temperature $\tilde T_b$ within each in. We define a `mean-subtracted' velocity reconstruction $\hat v_r(\bx)$ by:
\begin{equation}
    \hat{v}_r(\bx) = \sum_b\sum_{i\in b}W_i^v({\tilde T}(\btheta_i)-\tilde T_b)\delta^3(\bx-\bx_i)\,.
\end{equation}
This subtraction projects out the modes of $\hat v_r$ that vary slowly in redshift (including those that are constant across the sky, the radial analog of the integral constraint in galaxy clustering~\citep{deMattia:2019vdg}). 
In fact, in our analysis, we apply this mean subtraction to the galaxy fields as well.  Sampling the redshift distribution of randoms from data is generally understood to alter the clustering measurements~\citep{deMattia:2019vdg}, and our treatment is equivalent to including a radial integral constraint that accounts for this effect. 
In what follows, we will use $N_z=25$ redshift bins and apply the same mean subtraction to both our data and the surrogate fields of \S\ref{sec:surrogates}, for both $\hat v_r$ and galaxy fields.

We now model the remaining foreground contributions. At the small angular scales ($\ell>2000)$ the CMB contains a series of signals, including the primary CMB, the tSZ effect, the CIB:
\begin{equation}\label{eq:tobs_decomp}
T_{\rm obs}^\nu = T_{\rm prim} + T_{\rm kSZ} + T_{\rm tSZ}^\nu + T_{\rm CIB}^\nu+\ldots\,,
\end{equation}
where the tSZ, CIB terms depend on frequency. We approximate these contributions as biased tracers of the matter density field, $\bfg^\nu\, \delta_m$, where $\bfg^\nu$ is the `foreground bias' at frequency $\nu$. At each frequency, the mean of the velocity-reconstruction estimator of Eq.~\eqref{eq:est_definition} is therefore:
\begin{equation}\label{eq:hvr_with_bfg}
\langle \hat v_r^\nu(\bx)\rangle = {\underbrace{\bvel\, B(\bx)\, n_v(\bx)\, v_r^{\rm true}(\bx)}_{\rm kSZ}}+{\underbrace{\bfg^\nu\, W_{\rm CMB}(\btheta)\, n_v(\bx)\, \delta_m(\bx)}_{\rm foregrounds}} \,.
\end{equation}
In what follows, we focus on the tSZ and CIB contributions, the other contributions are sub-dominant compared to these on the scales we use in our analysis.\footnote{The linear integrated Sachs-Wolfe effect~\citep{1967ApJ...147...73S} is confined to the largest angular scales ($\ell < 100$), which our filter ($\ell > 2000$) removes. The nonlinear Rees-Sciama~\citep{1968Natur.217..511R} contributions are small compared to kSZ on these scales, and both weak-lensing and moving-lens~\citep{Birkenshaw:1983jhv,Hotinli:2018yyc} (now detected in \citet{Hotinli:2026mvl}) effects change sign with the lensing potential and transverse velocities, respectively, which are uncorrelated with the radial velocities. We use CMB maps with radio sources removed to reduce their contamination. Last, the primary CMB anisotropies also do not correlate with the galaxy density and only add reconstruction noise rather than biasing the estimator.}

The two terms in Eq.\ (\ref{eq:hvr_with_bfg}) depend differently on $\mu$, the cosine of the angle between the wavevector and the line of sight we defined in \S\ref{sec:pk_est}, so they enter different multipoles of the galaxy-velocity power spectrum. The density term enters the even multipoles, mostly the monopole $\ell=0$. Since $v_r^{\rm true} \propto \mu\,\delta_m$, the velocity term enters the odd multipoles, mostly the dipole $\ell=1$. In an idealized geometry, the two contributions separate cleanly, and the foreground does not bias the kSZ signal. The survey selection function (footprint, angular mask, completeness, and radial selection), however, introduces the usual window-function convolution that mixes multipoles, so that foreground power leaks from the monopole into the dipole and can contaminate the velocity signal. We will model this contamination with the surrogate methodology of \S\ref{sec:surrogates}. 

We model the foreground bias as:
\begin{equation}
    \bfg^{\nu} = A_{\rm tSZ}\, f_{\rm tSZ}(\nu) + A_{\rm CIB}\, f_{\rm CIB}(\nu)\,,
    \label{eq:bfg_sed_model}
\end{equation}
which is a more interpretable parameterization, rather than defining independent values for $b_{\rm fg}^\nu$ at each CMB map frequency. We model $f_{\rm tSZ}(\nu)$ assuming the analytic non-relativistic tSZ frequency spectrum~of~Ref.~\citep{Sunyaev1972}, and we model $f_{\rm CIB}(\nu)$ by a modified blackbody, for which we take a dust temperature $T_d=24\,$K and an emissivity index $\beta = 1.6$. We normalize these functions so that $f_{\rm tSZ}(150) = f_{\rm CIB}(150) = 1$. The amplitudes $A_{\rm tSZ}$ and $A_{\rm CIB}$ have a physical convention that is described in~\S\ref{sec:surrogates}. Throughout this work, we use the galaxy-velocity monopole $P_{gv}^{\ell=0}$ to fit these two amplitudes and to model their leakage into galaxy-velocity dipole and velocity-velocity correlations. 

\subsection{Galaxies and velocities in redshift space}

We measure galaxies and velocities in redshift space. At linear and mildly non-linear scales, the galaxy density in redshift space is related to linear matter density $\delta_m$ in real space by taking into account the Kaiser and Fingers-of-God effects. Following~\citet{Chaussidon:2026vmn}, we define:
\begin{equation}
    \delta_g^s(\bk)=\left(b_g+f_{\rm NL}\frac{2\delta_c(b_g-1)}{\alpha(k,z)}+f\mu^2\right)D_g(k,\mu,\Sigma_s)\,\delta_m(\bk)
\end{equation}
where $D_g(k,\mu,\Sigma_s)=1/(1+(k\mu\Sigma_s)^2/2)$ is a damping term that accounts for the Fingers-of-God effect, parametrized by the damping parameter $\Sigma_s$. This parametrization absorbs the nonlinearities, extending the validity of our modeling to mildly nonlinear conditions~\citep{Chaussidon:2026vmn}, as well as capturing observational effects due to redshift errors, for example.

On large scales, the radial velocity field is directly related to the matter density through the continuity relation, $v_r(k,\mu)=(ifaH\mu/k)\delta_m(k)$, and it thus provides an unbiased probe of large scale density fluctuations. In what follows, we approximate the radial velocity field in redshift space with a damping term similar to that of the galaxy, as:
\begin{equation}
    v_r^s(\bk)=\frac{ifaH\mu}{k}D_v(k,\Sigma_v) \delta_m(\bk)\,,
\end{equation}
where $D_v(k,\Sigma_v)={\rm sinc}(k\Sigma_v)$, parameterized by $\Sigma_v$. We will use these definitions when modeling the observed spectra using our surrogate fields framework.

\subsection{Surrogate fields}\label{sec:surrogates}\label{sec:surrogates_model}

In this section, we describe our `surrogate fields' framework. We use surrogate fields to model the galaxy and velocity auto- and cross-spectra. A surrogate field is a Monte-Carlo representation of a simplified random field that preserves the exact field-level covariance of a more complicated random field that includes survey geometry, light-cone evolution, photometric uncertainties, and so on. The surrogate fields framework allows us to model and calculate the expectation values for our power spectrum estimators and their covariance accurately on large scales $(k\gtrsim0.02\,{\rm Mpc}^{-1})$. A more complete introduction of the surrogate fields can be found in~\citet{Hotinli:2025tul}. Here, we give a brief overview. 

We start with a Gaussian field $\delta_{\rm lin}(\bx)$ in the bounding box. In order to model the galaxy field, we `paint' the values of $\delta_{\rm lin}(\bx)$ onto randoms as follows:
\begin{equation}
    S^{\rm sig}_g(\bx)\equiv\frac{N_g}{N_r}\sum_{j\in{\rm rand}}W_\beta\,\delta_G(\bx)\delta^3(\bx-\bx_j)\,,
\end{equation}
where
\begin{equation}
    \delta_G(\bk)=\left(b_g+f_{\rm NL}\frac{2\delta_c(b_g-1)}{\alpha(k,z)}+f\mu^2\right)D_g(k,\mu,\Sigma_s)\,\delta_{\rm lin}(\bk).
\end{equation}
We model the Poisson noise term by adding a Gaussian white noise; by defining an independent Gaussian random variable $\eta_j$ for each random $j$ as:
\begin{equation}
    S_g^{\rm noise}(\bx)\equiv s_n \frac{N_g}{N_r}\sum_{j\in{\rm rand}}W_j^g\,\eta_j\,\delta^3(\bx-\bx_j)\,,
\end{equation}
where $\eta$ satisfies:
\begin{equation}\langle\eta_j^2\rangle=\frac{N_r}{N_g}-\langle\delta_G(z_j)^2\rangle\,.
\end{equation}
The sum of the two terms gives our surrogate field $S_g(\bx)=S_g^{\rm sig}(\bx)+S_g^{\rm noise}(\bx)$. 

In each Monte-Carlo realization, we produce $S_g(\bx)$ and store its more optimized reparameterization as three separate fields: 
\begin{equation}
S_g(\bx)=b_g\,S_g^{(0)}(\bx)+f_{\rm NL}\,S_g^{(1)}(\bx)+s_n\,S_g^{(2)}(\bx)\,,    
\end{equation} 
as we fit $b_g$,~$f_{\rm NL}$~and~$s_n$ to data in~\S\ref{sec:results}. 

Modeling the radial velocity field follows a similar manner. We define the velocity surrogate field \textit{dipole} as: 
\begin{equation}\label{eq:Sv_decomp}
    S_{v}(\bx)=S_v^{\rm sig}(\bx)+S_{v}^{\rm fg}(\bx)+S_v^{\rm noise}(\bx)\,.
\end{equation}
We model the signal term as:
\begin{equation}\label{eq:surr_velo_dipole_sig}
    S_v^{\rm sig}\equiv b_v\, \frac{N_g}{N_r}\sum_{j\in{\rm rand}}W^v_j\,B(\bx_j)\,v_r^{\rm sim}(\bx_j)\delta^3(\bx-\bx_j)\,,
\end{equation}
where
\begin{equation}
    v_r^{\rm sim}(\bk)=\frac{ifaH\mu}{k}D_v(k,\Sigma_v)\delta_{\rm lin}(\bk)\,;
\end{equation}
and the foreground term as:
\begin{equation}\label{eq:surr_velo_dipole_fg}
    S_v^{\rm fg}\equiv b_{\rm fg}\,\frac{N_g}{N_r}\sum_{j\in{\rm rand}}W_j^v\,B(\bx_j)\,\delta_{\rm lin}(\bx)\delta^3(\bx-\bx_j)\,.
\end{equation}
For the noise term, we use a bootstrapping procedure based on real CMB maps.\footnote{Another approach would be to simulate the reconstruction noise by simulating random CMB realizations (including both the signal and noise) in each Monte Carlo realization. However this is not trivial as it requires modeling details such as the anisotropic noise and foreground correlations between different frequency maps.} In each Monte Carlo realization, we select a subset $S$ of the random catalog $\{1,\ldots,N_r\}$ of size $N_g$ (number of galaxies in the galaxy catalog) and define:
\begin{equation}
    S_v^{\rm noise}(\bx)\equiv s_{nv}\sum_{j\in{\rm rand}}M_j\, W_j^v\,\tilde T(\btheta_j)\, \delta^3(\bx-\bx_j)\,,\hspace{0.5cm}{\rm where}\hspace{0.5cm}M_j\equiv  \begin{cases} 
1 & \text{if } 1 \in S \\ 
0 & \text{if } 0 \notin S 
\end{cases}
\end{equation}
as we have done in~\citet{Hotinli:2025tul}. 

We parameterize $S_v$ in a way more optimized for our analysis:
\begin{equation}
     S_{v}(\bx) \equiv  b_v^\nu S_{v}^{(0)}(\bx) + b_{fg}^\nu S_{v}^{(1)}(\bx) + s_{nv}^\nu S_{v}^{(2)}(\bx)\,,
\end{equation}
and store these fields separately for each frequency $\nu$. Note here that because we include $B(\bx)$ — which is a constant value for our analysis choices — in our definition of the surrogate field, our fitted foreground bias $b_{\rm fg}^{\rm fit}$ relates to the $b_{\rm fg}^\nu$ introduced as $b_{\rm fg}\equiv B\times b_{\rm fg}^{\rm fit}$. We have maintained this choice made in~\citet{Chaussidon:2026vmn} for consistency. In following works, we will likely modify the definition for $S^{\rm fg}_\nu$ to avoid this redundancy. 

In what follows, in order to measure the foreground bias $b_{\rm fg}^{\nu}$, we calculate the galaxy-velocity \textit{monopole} $P^{\ell=0}_{gv}$ for each of our surrogate realizations. The surrogate velocity monopole is defined similarly to the dipole, with Fourier transforms set to spin-0 rather than spin-1 for calculating the galaxy-velocity correlation. This gives us the model prediction for the velocity-galaxy cross-correlation on the left-hand side of Eq.~\eqref{eq:Pgv_monopole} and allows us to measure and distinguish the tSZ and CIB contributions to our estimators.

For each Monte Carlo realization, we store the Fourier transforms of the $S_g$ and $S_v$ components, which we labeled with upper indices (0,1,2) above. This also allows us to fit the damping amplitudes $\Sigma_s$ and $\Sigma_v$, as we add the damping terms $D_g$ and $D_v$ directly to the multipoles after the window convolution. While we fit independent amplitudes for the biases $b_v$ and $b_{\rm fg}$, we use a common damping term $\Sigma_v$ for all frequencies, as $\Sigma_v$ should not depend on the choice of the CMB map.\footnote{Note, however, that as we will work with a photo-$z$ survey in this paper, the effect of damping will be much less pronounced in our case, since we do not have access to the smallest scales: photo-$z$ errors nullify our galaxy and velocity modes on scales well above where these effects become significant. We will not be able to measure these terms unambiguously (we can only put upper limits), but we will marginalize over their contribution, which in turn serves to make our analysis more robust.}  

Our pipeline also stores a term proportional to $f_{\rm NL}$ for monopole. In this work, we will not use the monopole signal, i.e., $P_{gv}^{\ell=0}(k)$, as a probe of PNG, since this term is not a direct measure of the kSZ signal. It probes the correlations between galaxies and foregrounds, such as tSZ and CIB, which we assume are biased tracers of galaxies; and may be subject to systematics or contributions that are beyond the scope of this paper for us to model. We nevertheless define $P_{gv}^{\ell=0}(k)$ monopole with such a term for consistency and to account for the leakage of foregrounds into velocity-galaxy dipole and velocity-velocity correlation signals. 

In what follows, we omit the use of the largest scales in our monopole measurements. We also do not fit $f_{\rm NL}$ to our monopoles in our analysis in \S{\ref{sec:results}}, where we indeed see a subtle hint of a $1/k^2$--like contribution within 1–2 sigma significance, although it is subdominant compared to foreground contributions we are fitting. We refrain from interpreting this as either systematics or PNG-like.

Lastly, in this work, we analyze a photometric survey, so we must extend the expressions we introduced above to incorporate the photometric redshift errors. We do so by defining for each galaxy the observed (photometric) and true redshifts, $z_{\rm obs}$ and $z_{\rm true}$, respectively, and $\sigma_z$, with which we denote the photo-$z$ errors. The galaxy catalogs generally contain $(z_{\rm obs},\sigma_z)$ values for each object, but $z_{\rm true}$ is unknown. Here, we deconvolve the observed ($z_{\rm obs},\sigma_z)$ values from the galaxy catalog in order to infer the underlying triple $(z_{\rm obs},\,z_{\rm true},\,\sigma_z)$ distribution. We define the comoving positions of galaxies at observed and true redshifts as $\bx_{\rm obs}=(\btheta,z_{\rm obs})$ and $\bx_{\rm true}=(\btheta,z_{\rm true})$, respectively. We then modify the definitions we made in this section by {placing the delta functions at observed locations}: $\delta^3(\bx-\bx_i)\rightarrow\delta^3(\bx-\bx^{\rm obs}_i)$. We place the $\delta_G$ and $v_r^{\rm sim}$ fields at \textit{true} locations. We also let the weights $W_i^g$ and $W_i^v$ be functions of both $z_{\rm obs}$ and $\sigma_z$. A more detailed description of our framework with photo-$z$ errors can be found in \citet{Hotinli:2025tul}.

\section{Data}\label{sec:data}

\begin{figure*}[tbp]
    \centering
    \includegraphics[width=\textwidth]{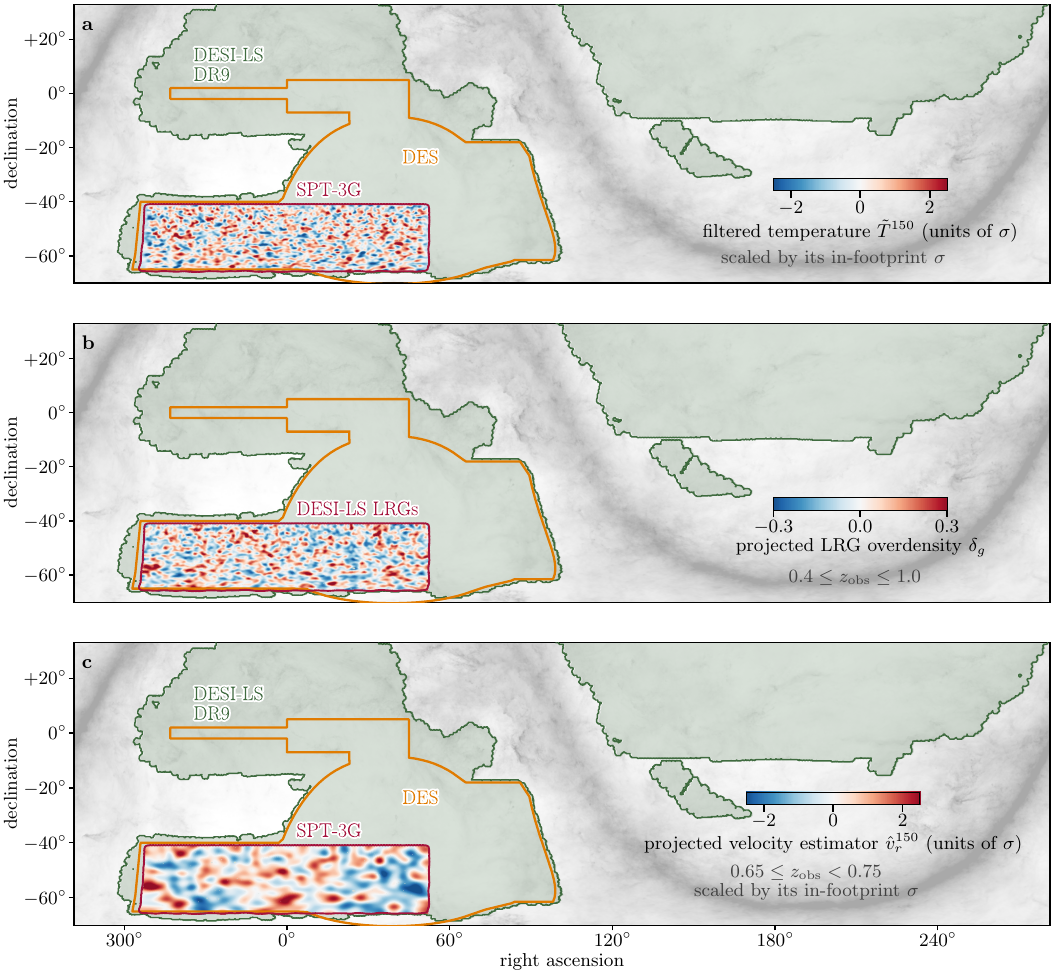}
    \vspace{-4pt}
    \caption{Filtered CMB temperature, projected LRG overdensity and reconstructed velocities over the SPT-3G footprint, shown in equatorial coordinates. \textbf{(a)} The filtered 150-GHz temperature evaluated at the main-sample galaxy positions, with the beam equalisation and filtering of~Eq.~\eqref{eq:beam_equalise}. For display, we smooth it with a Gaussian of $0.8^\circ$ full width at half maximum (FWHM), subtract its in-footprint median, and scale it by its standard deviation. \textbf{(b)}: The projected LRG overdensity over $0.4\leq z_{\rm obs}\leq1.1$ (using our \textit{galaxies-minus-randoms} prescription) and smoothed with the same Gaussian. \textbf{(c)} Projected 150-GHz radial-velocity estimator of the main sample over same footprint. We show the estimator in the shell $0.65\leq z_{\rm obs}<0.75$, calculated from the filtered temperatures as described in \S\ref{sec:vel_estimator}. The pixel are smoothed with a $2.5^\circ$ FWHM Gaussian. We then subtract the median of the estimator and scale the result by its standard deviation inside the footprint. The projected estimator contains reconstruction noise and residual foregrounds, and we assign no significance to individual structures. In all panels, the grey shading shows the \textit{Planck} GNILC 353-GHz dust intensity~\citep{PlanckDust2016}. Green shading shows the DESI-LS DR9 imaging footprint~\citep{Dey:2018}, traced by its randoms, and the orange curve marks the DES boundary~\citep{2005astro.ph.10346T}; neither used as a footprint in our analysis. }
    \label{fig:footprint}
\end{figure*}

In this section, we describe the CMB maps and the galaxy catalogs we use in this analysis. One novel aspect of this work is that for CMB, we use the maps and ancillary products of the SPT-3G D1 release \cite{SPT-3G:2026krg} (\S\ref{sec:data_cmb}). This is the first time SPT maps are used for kSZ velocity reconstruction analysis. Overall, we have found these products to be straight-forward to work with, and the maps we use to be of the highest quality for our purposes. For the galaxy catalog, we follow \citet{Hotinli:2025tul} and use DESI-LS DR9. We will transition to using the more recent DR11 dataset of the same survey in upcoming works. Figure~\ref{fig:footprint} shows various footprints of related surveys, DESI-LS DR9, DES, and SPT-3G. 

\begin{figure*}[!h]
    \centering
    \includegraphics[width=0.95\textwidth]{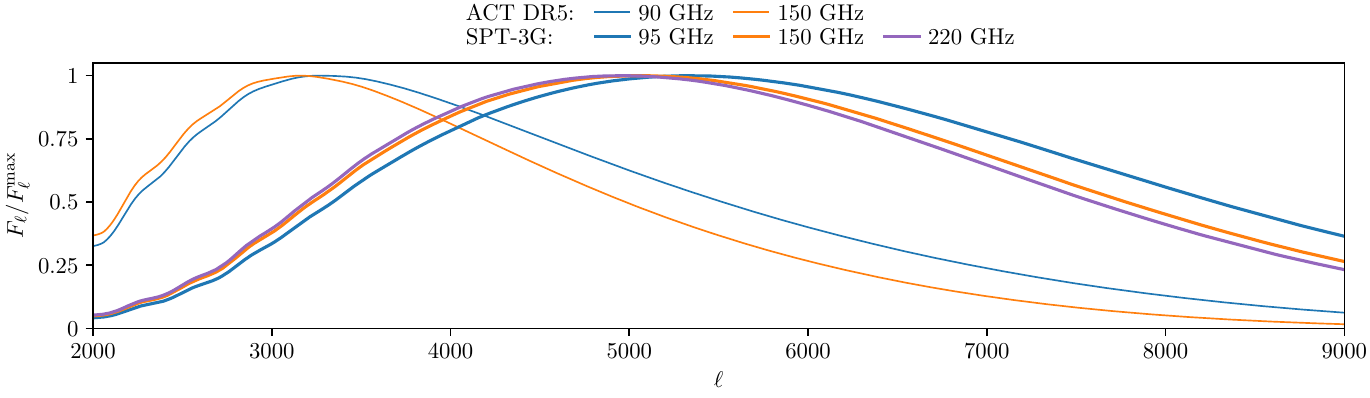}
    \vspace{-11pt}
    \caption{The CMB filters $F_\ell$ at each frequency, divided by their maximum $F^{\rm max}_\ell$, for ACT DR5 at 90 and 150 GHz from~\citet{Hotinli:2025tul}, and SPT-3G D1 at 95, 150 and 220 GHz used here. SPT-3G D1 filters peak at a higher $\ell$ around $\sim5100$ compared to ACT DR5 around $\ell\sim3600$.}
    \label{fig:filters}
\end{figure*}

\subsection{SPT-3G D1 maps and ancillary products}\label{sec:data_cmb}

In this analysis we use the full-depth coadded temperature maps at 95, 150, and 220\,GHz from the SPT-3G D1 release~\citep{SPT-3G:2026krg,SPT-3G:2025bzu}, made from two years (2019--2020) of observations of the SPT-3G Main field with the SPT-3G receiver~\citep{SPT-3G:2021vps} on the South Pole Telescope~\citep{Carlstrom:2009um}. The Main field spans declination $-42^\circ$ to $-70^\circ$ and right ascension $-50^\circ$ to $50^\circ$. The maps are stored in HEALPix~\citep{2005ApJ...622..759G} format at $N_{\rm side}=8192$, and we downgrade them to $N_{\rm side}=4096$ for processing.

For the analysis mask, we adopt the apodised borders-and-objects mask provided in the SPT-3G D1 release, which apodises both the survey boundary and the disks covering the high-signal-to-noise objects (point sources and clusters) that are masked in the SPT timestream high-pass filter. We build the CMB pixel weight by thresholding the per-frequency inverse-noise weight maps $w^\nu(\btheta)$ released with the data, each normalized to its maximum.
\begin{equation}\label{eq:wcmb}
W_{\rm CMB}(\btheta) = W_{\rm apo}(\btheta)\,\prod_{\nu}\mathbf{1}\big[w^\nu(\btheta) > 0.05\big]\,,
\end{equation}
where $W_{\rm apo}$ is the apodised borders-and-objects mask above, $\mathbf{1}[\cdot]$ is the indicator function and the product runs over the three frequencies. This threshold removes the noisy non-uniform-coverage boundary of the field. We use the resulting $W_{\rm CMB}$ at all three frequencies (the footprint corresponds to $f_{\rm sky} = 0.043$, $1780\,$deg$^2$).

For the beams $b_\ell^\nu$ we defined in \S\ref{sec:pipeline}, we use SPT-3G effective beams that are part of the auxiliary data products as described in \S5.2~of~Ref.~\citep{SPT-3G:2025rxd}~and~in~Refs.~\citep{SPT-3G:2026mcr,SPT-3G:2026krg}. These are defined as a product of the filter transfer function, a pixel window, and the SPT-3G D1 beam. As in Refs.~\citep{Hotinli:2026mvl,Chaussidon:2026vmn} we equalize the 150 and 220 GHz beams to the 95\,GHz beam, which is the broadest. We evaluate Eq.~(\ref{eq:fl_def}) at 95\,GHz with $C_\ell^{{\rm tot},\,95}$ fitted to the 95\,GHz CMB map; and rescale the filter at 150 and 220\,GHz by the beam ratio so that all three reconstructions share the same effective $\ell$-weighting:
\begin{equation}\label{eq:beam_equalise}
F_\ell^\nu = \frac{b_\ell^{\,95}}{b_\ell^{\,\nu}} \, F_\ell^{95}\,, \hspace{1.5cm} \nu \in \{150,\,220\}\,.
\end{equation}
While this choice makes us lose some SNR, it nevertheless simplifies our analysis by making the velocity bias of Eq.~(\ref{eq:bv_def}) the same at all three frequencies and sets the null maps ($90-150$,~$90-220$,~and~$150-220$) we describe in~\S\ref{sec:results_nulls} free of the kSZ signal. We show the filters used in this analysis in Figure~\ref{fig:filters}.

Below in \S\ref{sec:results_nulls}, we use the two half-depth coadds released with the SPT-3G D1 maps, formed by randomly partitioning the 2019–2020 dataset into two equal-depth halves for null tests. The release additionally provides 500 sign-flip noise coadds and 500 sets of mock-observation input and output maps. We do not use these in the present analysis since the surrogate framework provides our covariance estimate.

\subsection{DESI-LS LRG catalogue}\label{sec:data_gal}

As we stated above, for the galaxy survey, we use the DESI-LS DR9 LRGs~\citep{Dey:2018,DESI:2022gle,Zhou:2023gji}. DESI-LS combines data from three optical surveys (DECaLS, BASS, and MzLS) covering approximately $14000\,{\rm deg}^2$ of the extragalactic sky. We use the main and extended LRG samples of~\cite{Zhou:2023gji}. The main sample is the DESI LRG target sample~\citep{DESI:2022gle}. The extended sample uses looser selection cuts, which give 2--3 times the density and slightly larger photometric-redshift errors. The main sample is nested within the extended sample, with identical per-object photometric redshifts.

Whereas \citet{Hotinli:2025tul} restricted its ACT analysis to the North Galactic Cap (NGC), here we work in the Galactic South, where the SPT-3G Main field overlaps with DESI-LS. The southern LRG catalog is built from both the DECam imaging of DECaLS and the photometry of DES~\citep{DESI:2022gle,Zhou:2023gji}.

We restrict to the photometric redshift range $0.4 \le z_{\rm obs} \le 1.1$, apply the veto masks and quality cuts described in~Ref.~\citep{White:2021yvw} and \S3.3 of~Ref.~\citep{Zhou:2023gji}, and restrict to sky regions where $W_{\rm CMB}(\btheta) > 0$. We summarize the properties of our galaxy samples after these cuts in Table~\ref{tab:lrg_samples}. Throughout, number densities are measured over the approximately 1400~deg$^2$ that the galaxies actually occupy, as traced by the surviving randoms, rather than over the full 1780~deg$^2$ where $W_{\rm CMB}>0$.

\begin{table}[t!]
\centering
\caption{DESI-LS LRG sample properties in the SPT-3G D1 footprint. The main selection is nested within the extended one. The number densities are measured over the galaxy-occupied area traced by the randoms. The effective area and $f_{\rm sky}$ are evaluated where $W_{\rm CMB}>0$. We list the samples after the photometric-redshift-error cut described in \S\ref{sec:data_gal}. The photo-$z$ entries are the constant errors we adopt from~\citep{Zhou:2023gji}.}
\vspace{1.5mm}
\setlength{\tabcolsep}{8pt}
\begin{tabular}{lcc}
\toprule
Property & Main & Extended \\
\midrule
$N_g$ & 690{,}135 & 1{,}742{,}115 \\
$\bar n_g$ [Mpc$^{-3}$] & $1.39\times 10^{-4}$ & $3.50\times 10^{-4}$ \\
$z_{\rm min}$, $z_{\rm max}$, $z_*$ & 0.4, 1.1, 0.754 & 0.4, 1.1, 0.754 \\
effective area [deg$^2$] & $\approx 1780$ & $\approx 1780$ \\
$f_{\rm sky}$ & 0.043 & 0.043 \\
photo-$z$ scatter $\sigma_z/(1+z)$ & 0.024 & 0.027 \\
\bottomrule
\end{tabular}
\label{tab:lrg_samples}
\end{table}

Our analysis in this papers differs from \citet{Hotinli:2025tul} two notable ways. First, we remove the $5\%$ of galaxies with the largest fractional photometric-redshift uncertainty $\sigma_{z}/(1+z)$. This sets a threshold of $\sim0.056$ for the extended sample and $\sim0.055$ for the main sample. The removed galaxies have a median $\sigma_{z}/(1+z)$ of $0.071$ in both samples, compared to around $0.027$ (extended) and $0.024$ (main) for the galaxies we keep. Second, we adopt a constant $\sigma_z$ value for our galaxies and randoms in our analysis, which we set using the spectroscopically confirmed normalized median absolute deviation of the scatter of LRG photometric redshifts from Ref. \citep{Zhou:2023gji}. We use $\sigma_z = 0.027\,(1+z)$ (extended) and $0.024\,(1+z)$ (main). 

We weight every galaxy equally in the galaxy density and reconstructed velocity (i.e. $W_g=W_v$), and with our choice of constant photo-$z$ errors, these weights are the same for every galaxy. As in Ref.~\citep{Hotinli:2025tul}, we do not include an FKP weight~\citep{Feldman:1993ky}. As described in \S\ref{sec:surrogates}, we assign each random object a $(z_{\rm obs}, z_{\rm true})$ pair drawn from a deconvolution of the observed redshift distribution of the galaxy sample after with the constant scatter above. We apply the same redshift and quality cuts to the randoms as to the galaxy catalog.

\section{\texorpdfstring{Results: SPT-3G $\times$ DESI-LS}{Results: SPT-3G x DESI-LS}}\label{sec:results}

We now describe our analysis and results in this section. Our results contain a number of first detections: We detect the velocity-galaxy correlation dipole, monopole, and velocity-velocity autocorrelation for the first time with SPT-3G. We extend the kSZ analysis up to date to 220 GHz CMB maps for the first time, as we find that the SPT-3G maps at this frequency are more than adequate for this analysis, where we detect the galaxy-velocity dipole at high significance for both the main and extended catalogs, and we detect the galaxy-velocity monopole and velocity-auto at a significance larger than $3\sigma$ for the extended catalog. For the first time, we analyze the galaxy-velocity correlation together with the galaxy-galaxy spectrum, and provide joint constraints that improve $f_{\rm NL}$ measurement from galaxies alone. We start by introducing our statistical analysis choices in \S\ref{sec:statistical_analysis}, and describe our results in the rest of this section. 

\subsection{Statistical analysis}\label{sec:statistical_analysis}

\subsubsection{Parameter constraints}

In \S\ref{sec:pipeline} we introduced our pipeline, where we also defined the parameters we use to model our observables: galaxy-velocity correlation dipole and monopole, as well as velocity-velocity and galaxy-galaxy auto-correlations. These are the galaxy bias $b_g$, three velocity bias parameters $(b_v^{95},~b_v^{150},~b_v^{220})$ for each CMB frequency, the amplitudes that parametrize the foreground biases $(A_{\rm tSZ},~A_{\rm CIB})$, the local PNG amplitude $f_{\rm NL}$, a parameter that allows us to fit for the galaxy Poisson noise $s_n$, and three additional parameters for the velocity reconstruction noise $(s_{nv}^{95},~s_{nv}^{150},~s_{nv}^{220})$. We also defined the damping parameters $(\Sigma_g\,\Sigma_v)$ for redshift-space distortions. 

We infer these parameters by Monte Carlo Markov chain sampling of a Gaussian bandpower likelihood $\mathcal L$ that we define by:
\begin{equation}
    \log \mathcal L(\theta)=-\frac{1}{2}\log \det \mathcal C(\theta)-\frac{1}{2}(\boldsymbol{t}(\theta)-\boldsymbol{d})^T\mathcal{C}^{-1}(\boldsymbol{t}(\theta)-\boldsymbol{d})+\rm const\,,
\end{equation}
where $\mathbf{d}$ is a `data' vector that can contain all or some of the bandpowers, $P_{gv}^{\ell=1}(k)$, $P^{\ell=0}_{gv}(k)$, and $P_{vv}(k)$ at each frequency, and $P_{gg}(k)$. 
Here, $\mathbf{t}(\theta)$ is our `theory' predictions given (including all parameters for clarity) $\theta\in\{b_g\,,s_n,\,f_{\rm NL}\,, b_v^{95},\, b_v^{150},\, b_v^{220},\, A_{\rm tSZ},\, A_{\rm CIB}\, s_{nv}^{95}, ~s_{nv}^{150}, ~s_{nv}^{220},\,\Sigma_s,\,\Sigma_v\}$. The matrix $\mathcal C(\theta)$ is the bandpower-covariance we model using surrogate fields. Our surrogate field machinery computes all entries of $\mathcal{C}$, including correlations of bandpowers between different frequencies.

Later in this section, we define and use four likelihoods with different combinations of the spectra we measure and model:
\begin{enumerate}
    \item In \S\ref{sec:results_pgg}, we use the galaxy auto-power spectrum only, which is used to fit $b_g$ and $s_n$, and these are then held fixed when we measure the galaxy–velocity and velocity-auto signals separately.
    \item  In \S\ref{sec:results_monopole}, we fit foreground parameters using a likelihood including only the galaxy monopoles $P_{gv}^{\ell=0}$ in each frequency (also see below).
    \item In Sections~\S\ref{sec:results_dipole},~\S\ref{sec:results_pvv},~and~\S\ref{sec:results_fnl_vv_vg} we use the galaxy--velocity dipole $P_{gv}^{\ell=1}(k)$ together with the velocity auto-power monopole $P_{vv}^{\ell=0}(k)$.
    \item Finally in \S\ref{sec:results_fnl_vv_vg_gg} we add the galaxy auto-power spectrum to the third likelihood above, and use this for our final $f_{\rm NL}$ constraints.
\end{enumerate}

We make a few simplifying analysis choices: 
\begin{itemize}
\item We measure the galaxy-galaxy correlation separately to fit $b_g$ and $s_{n}$, which we use as fixed parameters (after marginalizing over $f_{\rm NL}$) for the analysis of the velocity reconstruction separately from $P_{gg}$. We fit these parameters together with the others only for our joint likelihood in~\S\ref{sec:results_fnl}. 
\item We fit the galaxy-velocity monopole by itself to infer the values of $A_{\rm tSZ}$ and $A_{\rm CIB}$ from the frequency dependence of Eq.~\eqref{eq:bfg_sed_model}. Here, we omit using the four largest-scale bandpowers to avoid interpreting the likely-systematic contribution at these scales, and set $f_{\rm NL}=\Sigma_v=0$. We fix $b_g$, $s_{n}$ and $\Sigma_s$ to their measured values from the $P_{gg}$ measurement. 
\item Although the surrogate framework lets us evaluate $\mathcal{C}(\theta)$ at any parameter point, we hold $\mathcal{C}$ fixed in both the likelihood minimization and the sampling, following~\cite{Chaussidon:2026vmn}.
We evaluate $\mathcal{C}(\theta)$ at the fitted values of all parameters, and repeat the MCMC fits with the updated values until no value changes by more than $0.05\sigma$. 
\end{itemize} 
For all parameters, we use uniform priors. In the velocity fits we use $\bvel^\nu\in(0,2)$, $s_{nv}^\nu\in(0.5,3)$, $\fnl\in(-500,500)$, and $\sigmav\in(0,50)$\,Mpc. In the monopole fits we sample the two foreground amplitudes as coefficients of the surrogate templates, with priors $(-0.01,0.01)$. In the fits that include the galaxy auto-spectrum, we use $b_g\in(1,4)$ and $\Sigma_s\in(0,50)$\,Mpc.

We sample the likelihood with \texttt{emcee}~\citep{ForemanMackey:2012ig}, seeding the walkers from a multi-start minimization of the likelihood.
We report constraints as marginalized posterior medians with $1\sigma$ uncertainties. For the goodness of fit, the signal-to-noise ratios, and the model predictions in the figures, we use the maximum-likelihood point of each fit.

\subsubsection{The $\chi^2$ statistics, detection SNR and null tests}

We assess whether our models are a good fit by computing the $\chi^2$ statistic separately for our observables:
\begin{equation}
    \chi^2=(\boldsymbol d-\boldsymbol m)^T\mathcal {C}^{-1}(\boldsymbol d-\boldsymbol m)\,,
\end{equation}
where $\boldsymbol m$ is the best-fit model and the covariance matrix $\mathcal C$ here is calculated at the best-fit model.

We quantify a detection signal-to-noise (SNR) value by:
\begin{equation}\label{eq:snr_def}
{\rm SNR} = \frac{\bd^T\,\mathcal{C}_0^{-1}\,\boldsymbol{m}_s}{\big(\boldsymbol{m}_s^T\,\mathcal{C}_0^{-1}\,\boldsymbol{m}_s\big)^{1/2}}\,,
\end{equation}
where $\boldsymbol{m}_s$ indicates the difference between our best-fit model and the model at the `null point', where the detected component is set to zero. For example, when showing the SNR for the dipole and monopole measurements below, we subtract our best-fit monopole and dipole leakage contamination from the signals, respectively, and we subtract both the reconstruction noise and the foreground contamination from the velocity auto when calculating its SNR. 

Similarly, the covariance $\mathcal{C}_0$ in our SNR is the covariance at the null point, where we subtract the best-fit signal from the covariance. For the velocity signal, for example, we set every $\bvel^\nu$ to zero and keep the foreground and noise terms in $\mathcal{C}_0$; for the foreground-induced monopole, we set $A_{\rm tSZ} = A_{\rm CIB} = 0$. We call Eq.~(\ref{eq:snr_def}) the detection SNRs, as in \citet{Hotinli:2025tul}, since $\mathcal{C}_0$ is the covariance under the null hypothesis; the SNRs we present in this work are these `detection' SNRs.

Joint SNRs over the three frequencies use the full covariance between them, and, for the dipole and the velocity auto-spectra, lie below the quadrature sum of the per-frequency SNRs mainly because the reconstruction noise is correlated between frequencies.
With $\mathcal{C}_0$ replaced by the covariance at the best fit, which includes the detected component, Eq.~(\ref{eq:snr_def}) gives the `amplitude' SNR.
The amplitude SNR measures how precisely the amplitude of $\boldsymbol{m}_{s}$ is determined, not how strongly the null hypothesis is rejected. Note the SNR values shown in~\citet{Chaussidon:2026vmn} for velocity auto are this `amplitude' SNR. 
Table~\ref{tab:main_consolidated} lists the detection and amplitude SNRs of the dipole, the monopole, and the velocity auto-spectra.

We report the goodness of fit as the reduced $\chi^2 / N_{\rm dof}$ and the probability-to-exceed ${\rm PTE}_{\rm gof} = \Pr(\chi^2 > \chi^2_{\rm obs})$. We quantify each null test by the probability $p_0$ of obtaining a larger $\chi^2$ than zero by chance and consider a fit or a test to pass if ${\rm PTE}_{\rm gof}$ or $p_0$ exceeds 0.05. 

We use frequency differences to test consistency between frequencies. The kSZ signal has the blackbody spectrum of the CMB, and the beam equalization of Eq.~(\ref{eq:beam_equalise}) gives the three frequencies the same response to it, so the kSZ dipole cancels in the difference between two frequencies. A contribution that differs between frequencies, such as foreground leakage, would not cancel. For each pair $(\nu_a, \nu_b)$ we test $\Pgv^{\nu_a, \ell=1} - \Pgv^{\nu_b, \ell=1}$ against zero, with the covariance of the difference from the surrogate ensemble at the best fit values, and quote $p_0$.

We also use half-splits from the SPT-3G half-depth coadds of \S\ref{sec:data}. We make two velocity reconstructions from these and test their difference against zero, the sky signal cancels in the difference, so a dipole there would come from the map noise.

\begin{figure}[tbp]
    \centering
    \includegraphics[width=\textwidth]{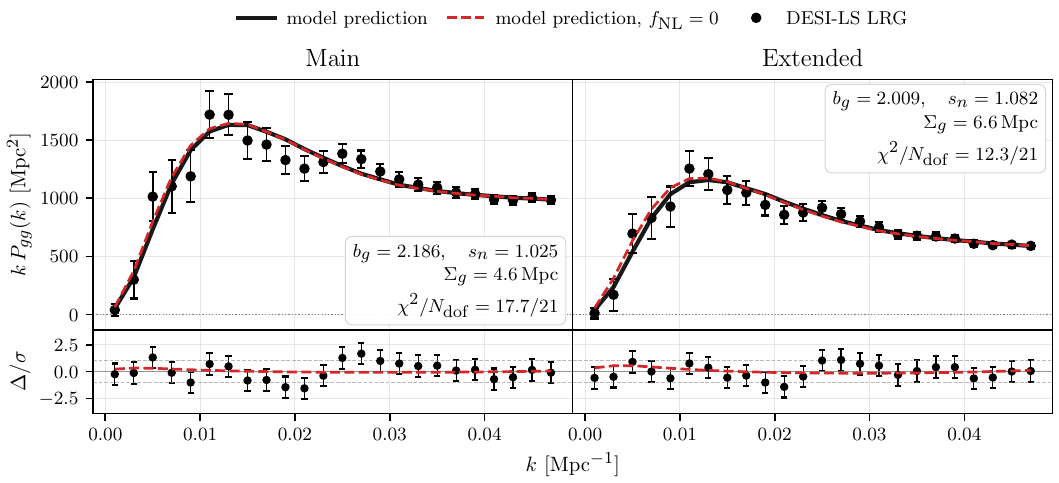}
    \caption{Galaxy auto-spectrum $\Pgg(k)$ measured within the SPT-3G D1 footprint for the DESI-LS LRG main and extended samples. Top sub-panels show the data with the best-fit model overlaid in black where we set $f_{\rm NL}$ free (red where we fix $f_{\rm NL}=0$); bottom sub-panels show residuals normalised by the diagonal covariance error. Insets show the best-fit $b_g$, $s_n$, $\Sigma_g$ values and $\chi^2/N_{\rm dof}$ where $f_{\rm NL}$ is set free and marginalized. Fixing $f_{\rm NL}=0$ gives best-fit values $b_g=2.156~(1.956)$ and $\Sigma_g=3.3~(4.8)$ for the main (extended) sample, with $\chi^2$ value increased by 0.22~(0.78). We use the model predictions with $f_{\rm NL}$ set free throughout the paper. The displayed goodness of fit is evaluated at these values and is acceptable in both samples, with ${\rm PTE}_{\rm gof}=0.67$ (main) and 0.93 (extended).}
    \label{fig:pgg}\label{fig:pgg_ext}
\end{figure}

\subsection{Galaxy auto-spectrum}\label{sec:results_pgg}

We begin our analysis by measuring galaxy auto-spectrum. We fit our prediction of $\hPgg(k)$ from our surrogate fields with $b_g$, $\fnl$ and the damping scale $\Sigma_g$ set free, and with the shot-noise amplitude $s_n$ at the Poisson level of the galaxies-minus-randoms field. For the main sample, the best fit is: 
\begin{equation}\label{eq:galaxy_auto_fnl}
    b_g=2.186^{+0.099}_{-0.087} \hspace{1.5cm} f_{\rm NL}=-33^{+42}_{-47}\,,
\end{equation}
$s_n=1.025$, and $\Sigma_g=4.6\,$Mpc, where $\chi^2/N_{\rm dof}=17.7/21$ and ${\rm PTE}_{\rm gof}=0.67$. The extended sample gives $b_g=2.009^{+0.089}_{-0.083}$, $s_n=1.082$ and $\Sigma_g=6.6\,$Mpc ($\chi^2/N_{\rm dof}=12.3/21$, ${\rm PTE}_{\rm gof}=0.93$) and $\fnl=-44^{+43}_{-45}$. We hold $b_g$, $s_n$ and $\Sigma_g$ fixed at these values in our analysis of $\Pgv$ and $\Pvv$ in what that follows, except in the joint fits of \S\ref{sec:results_fnl}. Figure~\ref{fig:pgg} shows the bandpowers and the best-fit model. Table~\ref{tab:main_consolidated} collects the parameters for both samples.

It is worth briefly discussing the $\fnl$ constraints we obtain from the galaxy auto-spectrum alone, which are worse but comparable to the DESI-LS analysis of \citet{Rezaie:2023lvi}, which uses the entire footprint of this survey. Here, we only use a fraction of the DESI-LS galaxies on the SPT-3G D1 footprint. Our analysis differs from such clustering analyzes in multiple ways, which may help explain our results. First, we are using a footprint (the SPT-3G footprint in Figure~\ref{fig:footprint}) that avoids the mixing of different galaxy and survey samples, so that we avoid using image-systematics treatments that cost sensitivity to these analyzes. Next, we do a 3-d analysis instead of a 2-d (purely angular) analysis in~Ref.~\citep{Rezaie:2023lvi}, which gains us more modes. Our pipeline also includes a mean-subtraction for the galaxy field (in addition to our velocity estimates) for both data and surrogate fields, which is essentially equivalent to including a radial integral constraint (RIC) that accounts for the corresponding loss of $\fnl$ sensitivity.

\subsection{Galaxy--velocity and velocity power spectra}\label{sec:results_spectra}

Next, we measure three spectra: the galaxy–velocity dipole, which carries the kSZ signal; the foreground-induced galaxy–velocity monopole; and the velocity auto-spectrum, which contains the kSZ signal, the reconstruction noise, and a foreground term. In this section, we discuss our detection of these signals and our measurement of the velocity and foreground biases. We describe our constraints on $f_{\rm NL}$ in \S\ref{sec:results_fnl}.

\subsubsection{Galaxy--velocity monopole $\Pgv^{\ell=0}$ and foreground decomposition.}
\label{sec:results_monopole}

\begin{figure*}[tb]
    \centering
    \includegraphics[width=\textwidth,trim={0 8 0 16},clip]{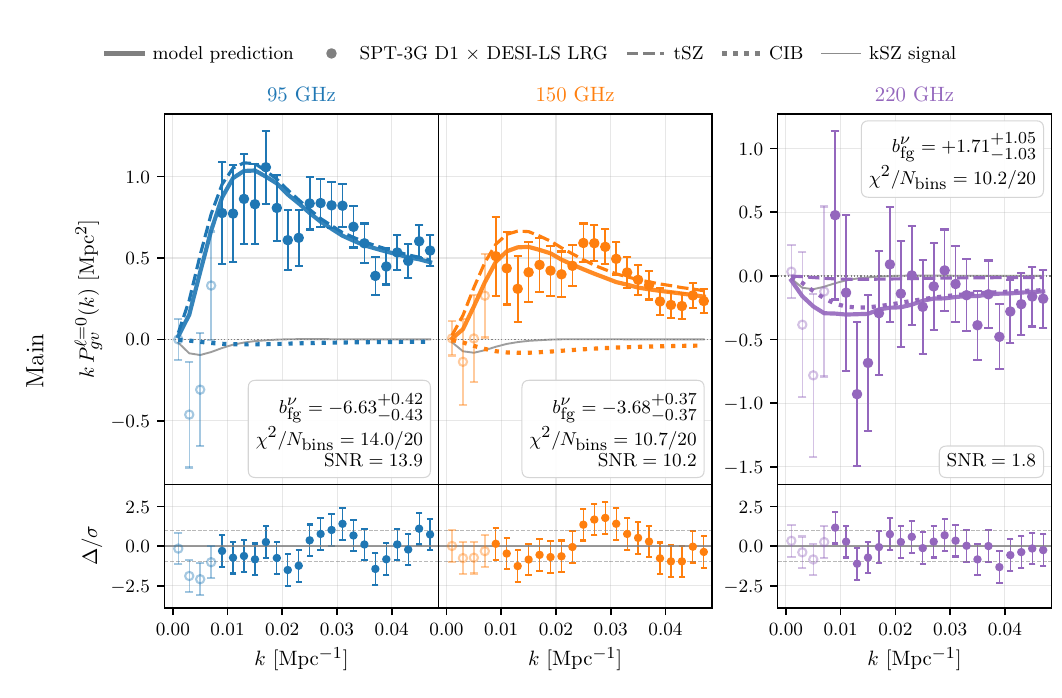}
    \caption{Per-frequency foreground-induced $\Pgv^{\ell=0}$ monopole within the SPT-3G D1 footprint for the DESI-LS LRG main sample.
Top sub-panels show $k\,\Pgv^{\ell=0}(k)$ with the model of the monopole fit and its tSZ, CIB and kSZ-signal components (thick dashed and dotted for tSZ and CIB, and thin solid line for kSZ). Bottom sub-panels show normalised residuals.
We fit the model omitting the first four bandpowers, which drawn as faded open markers.
Insets shows $\bfg$ values in units of $10^{3}\,\mu{\rm K}^{-1}$ (posterior median with the offsets to the $84$th and $16$th percentiles), $\chi^2/N_{\rm bins}$ and the foreground-monopole SNR of Eq.\ (\ref{eq:snr_def}) of that frequency.
Over these 20 bandpowers, the three monopoles give ${\rm PTE}_{\rm gof}=0.83$, 0.95 and 0.96 at 95, 150 and 220\,GHz.
The frequency dependence follows the tSZ+CIB model of Eq.\ (\ref{eq:bfg_sed_model}).}
    \label{fig:monopole_primary}
\end{figure*}

We start by measuring the galaxy-velocity monopole $\Pgv^{\ell=0}$ in order to quantify the foreground bias and its contamination of the galaxy-velocity dipole, which we will measure next. We detect the monopole signal at SNR $16.8$ over the three frequencies for the main sample, the monopole fit gives:
\begin{align}
    A_{\rm tSZ} &= (-4.18^{+0.28}_{-0.28})\times10^{3}\,\mu{\rm K}^{-1}, &
    A_{\rm CIB} &= (+0.51^{+0.34}_{-0.34})\times10^{3}\,\mu{\rm K}^{-1}, \label{eq:sed_amp_main}
\end{align}
which set the per-frequency foreground biases through Eq.~(\ref{eq:bfg_sed_model}), with posterior medians for the main sample:
\begin{equation}
    \bfg=\begin{cases}
        -6.6~{\rm at~95\,GHz}\\
        -3.7~{\rm at~150\,GHz}\,~\,\times10^3\mu{\rm K}^{-1}\,.\\
        +1.7~{\rm at~220\,GHz}
    \end{cases}
\end{equation}
Figure~\ref{fig:monopole_primary} shows the galaxy-velocity monopole bandpowers for the main sample, and surrogate fields model predictions at the best-fit values. The insets shows the per-frequency SNRs.
While measurement significance for the CIB  contribution is low for the main sample, we detect the CIB amplitude at an SNR exceeding 2 with the extended sample, for which we detect the monopole contribution at SNR value of $18.9$. We defer the majority of our results for the extended sample to Appendix~\ref{app:extended} for brevity.  

The foreground biases are negative at 95 and 150\,GHz, where the tSZ effect is a decrement, and positive at 220\,GHz, near the frequency where tSZ effect vanishes and the CIB dominates. The two-component model of Eq.~(\ref{eq:bfg_sed_model}) describes the monopole at every frequency, as can be seen from Figure~\ref{fig:monopole_primary}, so the measured frequency dependence agrees with that of the tSZ and CIB effects. We also show the subdominant contribution to the monopole from the kSZ effect with gray solid lines, induced by the window convolution.

\subsubsection{Galaxy--velocity dipole $\Pgv^{\ell=1}$.}\label{sec:results_dipole}

\begin{figure*}[tbp]
    \centering
    \includegraphics[width=\textwidth,trim={0 8 0 16},clip]{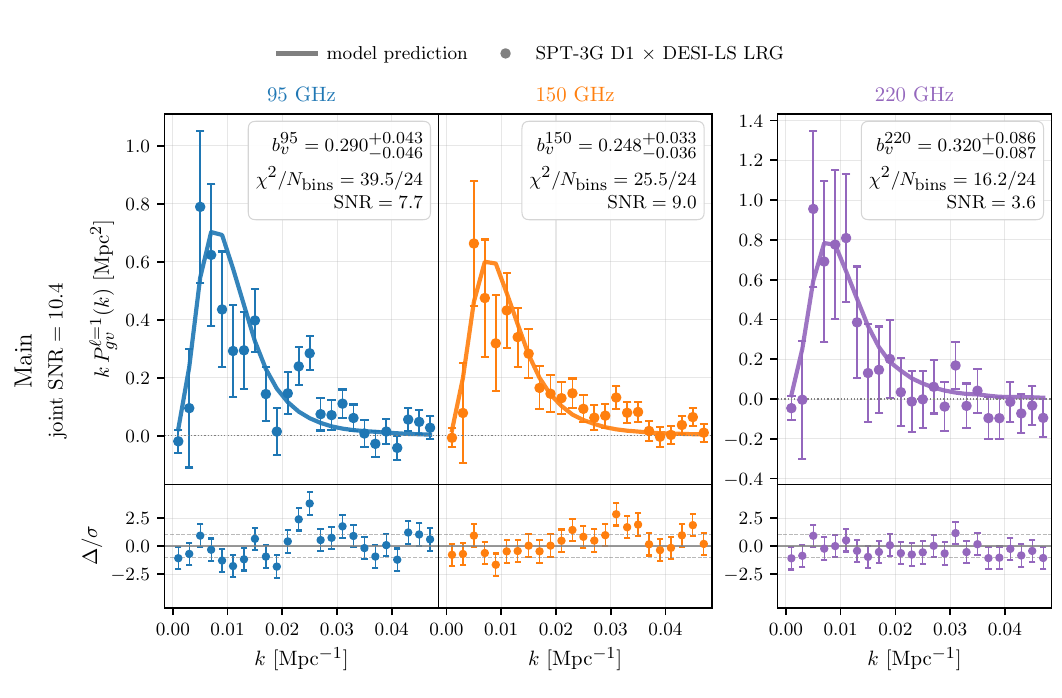}
    \vspace*{-24pt}
    \caption{Per-frequency $\Pgv^{\ell=1}$ dipole, which carries the kSZ signal, within the SPT-3G D1 footprint for the DESI-LS LRG main sample; columns are 95, 150, and 220\,GHz. Top sub-panels show $k\,\Pgv^{\ell=1}(k)$ with the model predictions. Bottom sub-panels show residuals normalised by the diagonal covariance error. Insets show $\bvel$ best-fits (posterior median with the offsets to the $84$th and $16$th percentiles), $\chi^2/N_{\rm bins}$ and the SNR of Eq.\ (\ref{eq:snr_def}). The joint SNR beside the row label is evaluated over all 3x24 dipole bandpowers. The 95\,GHz bandpower at $0.024<k<0.026\,{\rm Mpc}^{-1}$ drives the low goodness of fit discussed in \S\ref{sec:results_spectra}.}
    \label{fig:dipole_primary}
\end{figure*}

\begin{figure}[tbp]
    \centering
    \includegraphics[width=\textwidth]{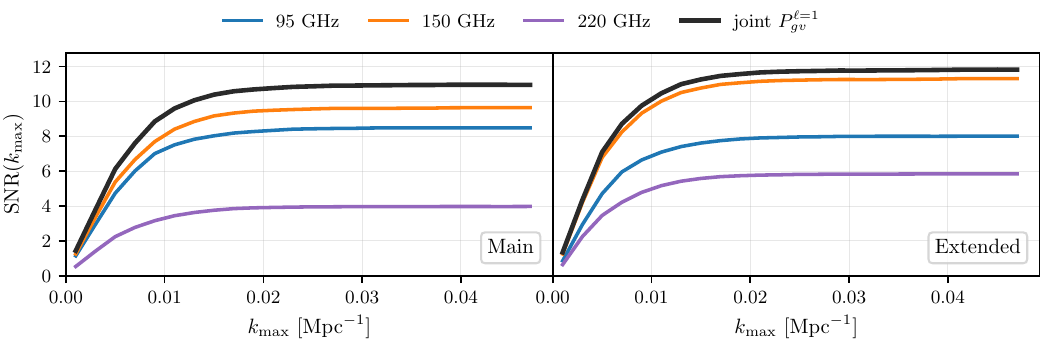}
    \vspace*{-20pt}
    \caption{Cumulative detection SNR of the main kSZ galaxy-velocity dipole $\Pgv^{\ell=1}$ for main and extended catalogs. The joint curve uses all three frequencies with the full covariance.}
    \label{fig:snr_kmax}
\end{figure}

Our main observable is the kSZ galaxy-velocity correlation dipole. We detect the kSZ dipole of the main sample at total SNR $10.4$. For each frequency, we measure velocity-galaxy dipole at:
\begin{equation}
    {\rm SNR}=\begin{cases}
        7.7~\rm at~95\,GHz,\\
        9.0~\rm at~150\,GHz,\\
        3.6~\rm at~220\,GHz.
    \end{cases}
\end{equation} 
This is the first detection of the galaxy-velocity correlation from CMB at 220 GHz. 
Figures~\ref{fig:dipole_primary} and \ref{fig:snr_kmax} show the bandpowers and the cumulative signal-to-noise of the velocity-galaxy dipole.

Having set the beam equalization as in Eq.~(\ref{eq:beam_equalise}); we expect the blackbody kSZ signal to give the same velocity bias at all three frequencies. We find that the data agrees, as the velocity fit with a per-frequency $\bvel$ gives:
\begin{equation}\label{eq:joint_bv_ext}
    \bvel=
    \begin{cases} 
        0.290^{+0.043}_{-0.046}~{\rm~at~95\,GHz}, \\ 
        0.248^{+0.033}_{-0.036}~{\rm~at~150\,GHz}, \\ 
        0.320^{+0.086}_{-0.087}~{\rm~at~220\,GHz},
\end{cases}
\end{equation}
for the main catalog. 

We also define a single $b_v$ value shared between all frequencies, and parameterize our surrogate field model predictions accordingly. Setting a single $\bvel$ raises $\chi^2$ by $3.0$ for two fewer parameters (probability $0.22$) and gives
\begin{equation}\label{eq:bottom_line_bv_pgv}
    \bvel=0.255^{+0.033}_{-0.035}.
\end{equation}

As we described above, the value of $b_v$ depends on the ratio between the true galaxy-electron power spectrum $P_{ge}^{\rm true}(k)$; and the fiducial $P_{ge}^{\rm fid}(k)$ we choose in our analysis. In this analysis, we use the halo model prescription in the \texttt{hmvec} code~\citep{Smith:2018bpn} and compute the $P_{ge}(k,z)$ at our central redshift $z_\star=0.754$. We used the Battaglia gas profile normalized in such a way that the enclosed gas mass equals the cosmic baryon fraction at the virial radius, as in Refs.~\citep{Hotinli:2025tul, Chaussidon:2026vmn}. This returns a gas profile that do not capture the baryonic feedback that is understood to reduce the ionized gas content of galaxy groups. Measurements find that ionized gas residing within the virial radius of these halos amounts to around only 25-30 percent of the cosmic baryon fraction~\citep[see e.g.][]{Hadzhiyska:2025mvt,Qu:2026zyh,Hotinli:2026zne}. 
Our measurement naturally returns a value lower than unity, in agreement with these observations. 

\subsubsection{Velocity auto-spectrum $\Pvv^{\nu\times\nu}$.}\label{sec:results_pvv}

In addition to the galaxy-velocity correlation, we also detect the kSZ signal in the three auto-spectra of reconstructed velocities. Our detection SNR for the three spectra is at 7.2 (8.1) for the main (extended) sample, the highest detection of the auto-correlations to date. The `amplitude' SNR (which we defined in \S\ref{sec:statistical_analysis}) is $\sim3.3$ for both samples.

The velocity auto-spectrum contains the kSZ signal, the reconstruction noise, and the foregrounds. For the detection SNR, we fit and subtract foreground and noise contributions from the velocity signal. We use the best-fit signal component only in the auto-spectra and include the foregrounds and noise in the covariance $\mathcal C_0$. 

\begin{figure*}[tbp]
    \centering
    \includegraphics[width=\textwidth,trim={0 8 0 16},clip]{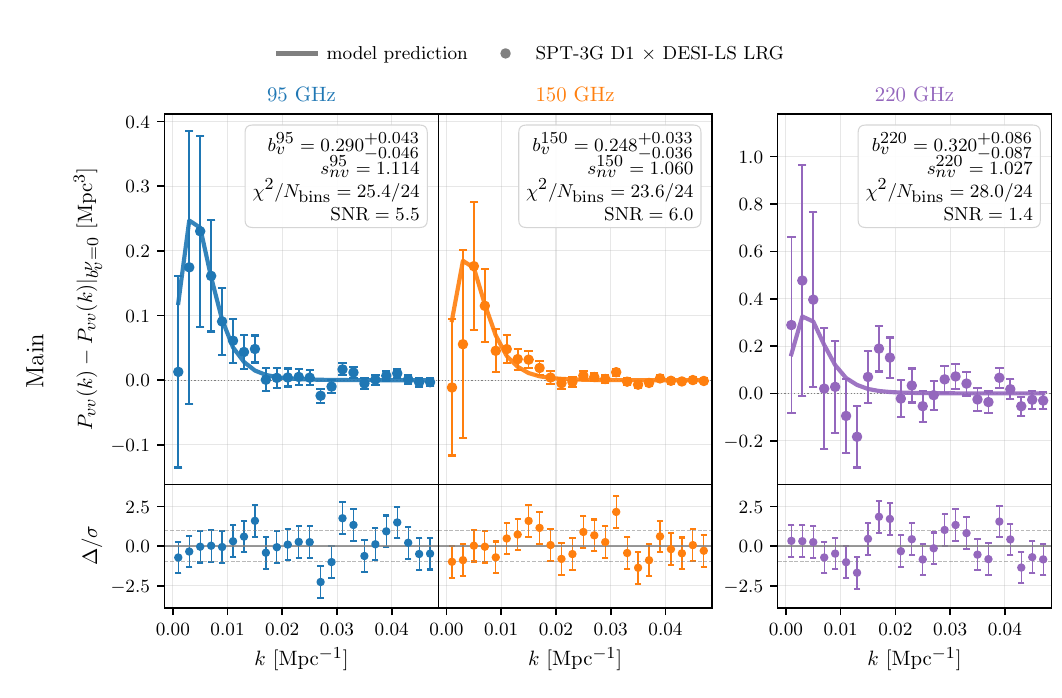}
    \vspace*{-24pt}
    \caption{Per-frequency velocity auto-spectrum within the SPT-3G D1 footprint for the DESI-LS LRG main sample. Top sub-panels show $\Pvv$ at each frequency minus the terms of Eq.~(\ref{eq:Sv_decomp}) that do not depend on $\bvel$ (i.e. the reconstruction noise and foreground contribution). Solid lines are our model prediction for the best-fit kSZ signal. Bottom sub-panels show normalised residuals. Insets show $\bvel^\nu$ values (posterior median with the offsets to the $84$th and $16$th percentiles), the posterior median of $s_{nv}^\nu$, $\chi^2/N_{\rm bins}$ and the SNR of Eq.\ (\ref{eq:snr_def}). The three auto-spectra give ${\rm PTE}_{\rm gof}=0.39$, 0.49 and 0.26 at 95, 150 and 220\,GHz. Jointly, the auto-spectra reach kSZ SNR $7.2$.}
    \label{fig:pvv_primary}
\end{figure*}

The SNR on the amplitude is noticeably lower than the detection SNR because the 95 GHz and 150 GHz auto-spectra are signal dominated on the largest scales, where most of the detection SNR arises. Figure~\ref{fig:pvv_primary} shows the bandpowers with the noise and foreground terms subtracted, and the corresponding best-fit signal we model.

The fitted reconstruction-noise amplitudes are: 
\begin{equation}
    s_{nv}=\begin{cases}
        1.11~{\rm at~95\,GHz}\,,\\
        1.06~{\rm at~150\,GHz}\,,\\
        1.03~{\rm at~220\,GHz}\,.        
    \end{cases}
\end{equation}
Each has uncertainty $0.01$, so the bootstrap slightly underestimates the reconstruction noise. We evaluate the covariance at the fitted amplitudes, which carries this calibration into the uncertainties.

\subsubsection{Goodness of fit.}
\label{sec:results_primary}
Our fits describe the data well, with one exception. All our fits pass the goodness-of-fit test (Table~\ref{tab:main_consolidated}), except for the overall velocity fit of the main sample (${\rm PTE}_{\rm gof}=0.02$). We find that this is mainly sourced by a single 95\,GHz galaxy-velocity dipole bandpower, at $0.024<k<0.026\,{\rm Mpc}^{-1}$, visible in Figure~\ref{fig:dipole_primary}, which lies $3.8\sigma$ above our best-fit model. This excess appears at no other frequency and in no other spectrum. We do not explore this systematic contribution further in this paper, but will do so in upcoming works. 

\subsection{Constraints on $f_{\rm NL}$}\label{sec:results_fnl}

Last on our parameter fits, we discuss our $f_{\rm NL}$ constraints. Local PNG enters our model only through the scale-dependent galaxy bias. The $\fnl$ enters the galaxy auto-spectrum through the square of this bias and the galaxy-velocity correlation linearly. 

\subsubsection{Constraints on $f_{\rm NL}$ from velocity-galaxy correlations}\label{sec:results_fnl_vv_vg}

We first discuss our constraints on $\fnl$ from velocity reconstruction alone, without the galaxy auto-spectrum. While the velocity auto-spectrum does not depend on $\fnl$, we find that jointly fitting the galaxy-velocity dipole and velocity-velocity autocorrelation improves $\fnl$ constraints compared to the galaxy-velocity dipole alone, as was seen in \citet{Chaussidon:2026vmn}, by reducing the degeneracy between $b_v$ and $\fnl$. Throughout this work, we will quote constraints on $\fnl$ from the the joint analysis of the galaxy-velocity dipole and velocity auto-spectrum.    

We sample $\fnl$ with the other free parameters of the velocity. Using the galaxy-velocity dipole and velocity-velocity correlation, we find
\begin{equation}\label{eq:joint_fnl}
    \fnl=-38^{+61}_{-61}\,,
\end{equation}
for the main sample. Setting $b_v$ to be shared gives a similar constraint at $\fnl=-35^{+62}_{-63}$ with a single $\bvel$. The extended sample gives a similar constraint at $\fnl=-50^{+63}_{-63}$, all consistent with zero and with the galaxy auto-spectrum at Eq.~\eqref{eq:galaxy_auto_fnl}. 

We test our $\fnl$ constraints a range of analysis choices: dropping the 95\,GHz outlier bandpower; keeping only the dipole at $k<0.02\,{\rm Mpc}^{-1}$; or fitting the monopole together with the dipole and velocity auto (recall that we fit the monopole separately in our baseline analysis). These choices move $\fnl$ by less than $0.4\sigma$ in both main and extended samples and do not change our constraints. 

These are the first measurements of PNG from kSZ using SPT-3G maps, on a relatively small footprint ($f_{\rm sky}\sim0.05$) in the sky, and from data collected between 2019--2020. We find SPT-3G maps give a constraint on $f_{\rm NL}$ comparable to our analysis on ACT~DR6$\times$DESI-LS in~\citet{Chaussidon:2026vmn}, which found the LRG sample gives $\sigma(f_{\rm NL})\simeq\pm44$ on a larger sky area. While our $\fnl$ constraints are noticeably weaker than those from the CMB bispectrum and from DESI galaxy clustering~\citep{Planck:2019kim, Chaussidon:2024qni} upcoming CMB and large-scale structure surveys will promptly close this gap. Our measurements show how largely insensitive the galaxy-velocity cross-correlation is to imaging systematics, which can add spurious power to galaxy clustering on the large scales where the PNG signal is largest~\citep{Chaussidon2022, Rezaie:2023lvi}, adding further value to kSZ velocity reconstruction and $\fnl$ measurements from these signatures. 

\subsubsection{Joint measurements of $f_{\rm NL}$ from velocity-galaxy and galaxy-galaxy auto}\label{sec:results_fnl_vv_vg_gg}

Next, we add the galaxy auto-spectrum to the velocity measurements. Combining galaxy auto with galaxy-velocity dipole, in principle, allows sample-variance cancellation between the two tracers~\citep{Seljak:2008xr,Munchmeyer2018} and measuring $\fnl$ without the limitation of cosmic variance. Our measurement significance here is not yet sufficient for us to be in this regime. 

\begin{figure*}[tbp]
    \centering
    \includegraphics[width=0.78\textwidth,trim={0 16 0 4},clip]{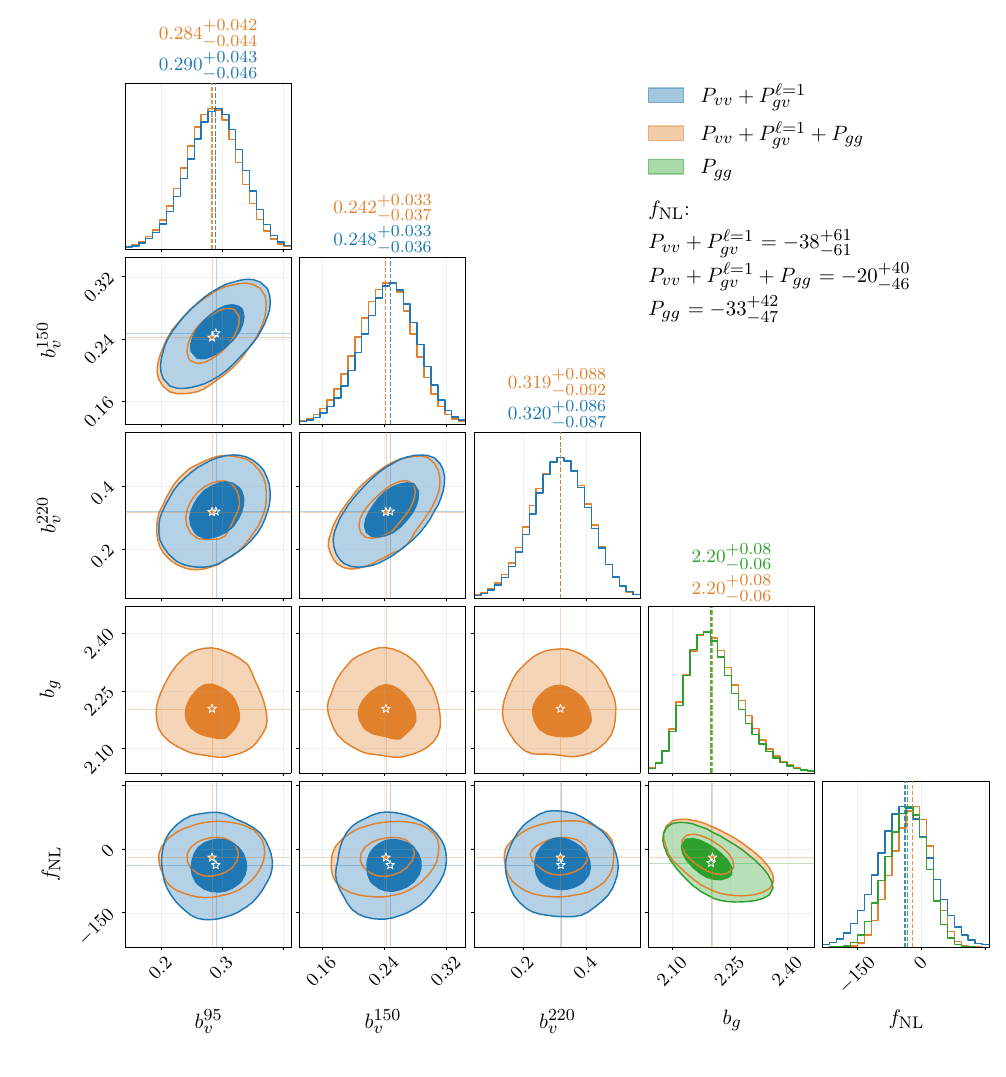}
    \vspace*{-12pt}
    \caption{The marginalised posteriors of $\fnl$, $b_g$ and $b_v$ values (the other parameters are marginalised) and $A_{\rm tSZ}$ and $A_{\rm CIB}$ are held at the maximum-likelihood point of the monopole fit in both fits. The velocity fit of $\Pvv$ and the $\Pgv$ dipole (blue) in \S\ref{sec:results_fnl_vv_vg} and the full fit in \S\ref{sec:results_fnl_vv_vg_gg}, which adds $\Pgg$ (orange), are overlaid. The $\fnl$ constraints of the two fits are $-38^{+61}_{-61}$ and $-20^{+40}_{-46}$, respectively, both consistent with zero. The green contour shows the constraints from $\Pgg$ alone. In the velocity fit in blue, the $b_g$ value is fixed at the $\Pgg$ best fit of \S\ref{sec:results_pgg}. The contours enclose $39\%$ and $86\%$ of the two-dimensional posterior mass; stars and dashed lines mark the posterior medians.}
    \label{fig:corner_pgg_full}
\end{figure*}

We free $b_g$ and $\Sigma_g$ and append the $24$ galaxy auto-spectrum bandpowers to the velocity data vector, with the $\Pgg\times\Pgv$ and $\Pgg\times\Pvv$ cross-covariances included. 
We hold $\bfg^\nu$ fixed at its best-fit value from the monopole measurements, and make all other choices as in \S\ref{sec:statistical_analysis}.
The full fit gives
\begin{equation}\label{eq:full_fnl}
    \fnl=-20^{+40}_{-46}\,,
\end{equation}
for the main catalog, where setting a single $\bvel$ for all frequencies gives a similar $\fnl=-17^{+40}_{-46}$, both consistent with zero. Adding the velocity spectra to the galaxy auto-spectrum moves $\fnl$ from $-33^{+42}_{-47}$ and narrows its interval by a few percent. 
The full fit gives $b_g=2.20^{+0.08}_{-0.06}$ with  $\bvel$ values consistent with the $\Pgg$ and the velocity reconstruction best-fit values. The velocity biases, noise amplitudes, and velocity damping scale move by at most $0.18\sigma$ from the velocity fits. Our constraints on these parameters are shown in Figure~\ref{fig:corner_pgg_full}.

It is valuable to note that the Fisher forecast of Ref.~\citep{Munchmeyer2018}, evaluated for our survey (volume, $k$ range, galaxy density, reconstruction noise, and photometric-redshift errors) with our measured $b_g$ and $b_v$, predicts a $\sim10\!-\!20\%$ smaller $\sigma(\fnl)$ than the galaxy auto-spectrum alone. In this first measurement, adding the velocities narrows the $\fnl$ interval by $\sim3\%$ in the full fit. We trace this shortfall to the correlation between the galaxy and velocity signals in our bandpowers. Fisher forecasts treat each Fourier mode as measured on its own, so on linear scales, the two signals are perfectly correlated. In a survey of our size, the window and the photometric-redshift errors mix the largest scale modes, and at $k\lesssim0.01\,{\rm Mpc}^{-1}$, where most of our $\fnl$ information lies, the correlation falls from $0.9$ to $0.4$ toward the largest scales. With the measured correlation and velocity amplitude, the same forecast predicts a $\sim4\%$ improvement, in line with what we find. The gap, therefore, reflects how the largest scales are treated at the current level of precision. In upcoming works, we will aim to improve our treatment of the largest scales to achieve better agreement between the estimator and a Fisher forecast.

\input{table_main.tex}

\subsection{Consistency and null tests}\label{sec:results_nulls}

\input{table_nulls.tex}

Finally, we discuss our consistency checks and null tests in this section. Our null tests consist of differencing our galaxy-velocity dipole correlation measurements at different frequencies. We also include half-split null tests, using SPT-3G half-depth coadds. 

\subsubsection{Frequency-difference nulls}

The kSZ signal has a black-body frequency dependence. Our treatment of the velocity-reconstruction estimator also assumes this: our measured velocity-galaxy correlations should be frequency independent. Our frequency-difference null tests this directly. 

We define three frequency pairs with $(\nu_\alpha,\nu_\beta)$ and measure the difference $P_{gv}^{\ell=1,\nu_\alpha}-P_{gv}^{\ell=1,\nu_\beta}$ using the corresponding covariances calculated with the surrogate fields, which share the density realization and reconstruction noise between the three frequencies.

\begin{figure*}[tbp]
    \centering
    \includegraphics[width=\textwidth,trim={0 4 0 11},clip]{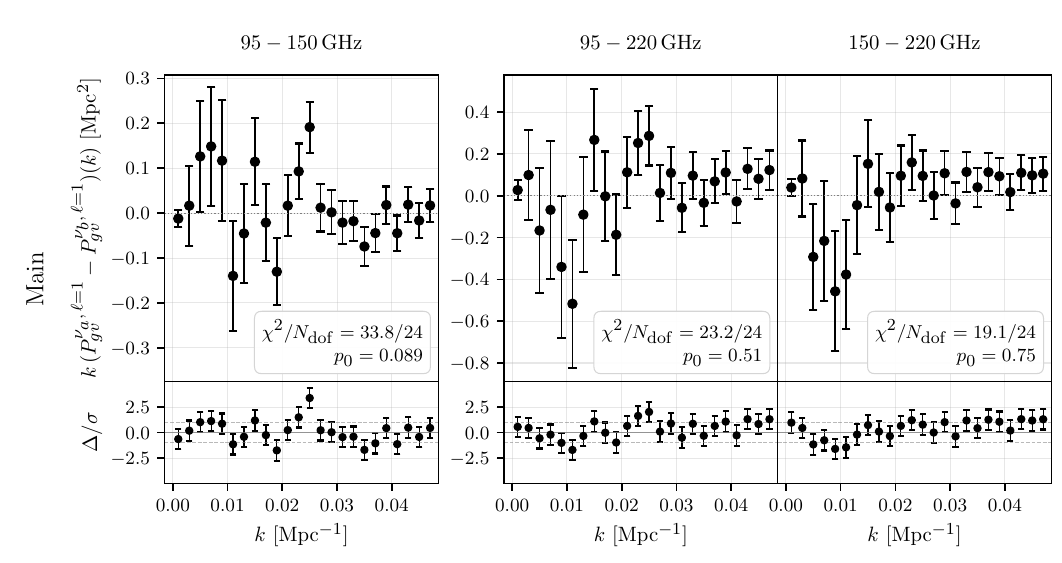}
    \vspace*{-24pt}
    \caption{Frequency-difference tests of the $\Pgv^{\ell=1}$ dipole for the DESI-LS LRG main sample. Columns show 95--150, 95--220, and 150--220\,GHz; bottom sub-panels show the corresponding normalised residuals. Error bars are the square roots of the diagonal of the difference covariance. Insets shows $\chi^2/N_{\rm dof}$ and the probability $p_0$ of Table~\ref{tab:nulls}. All three tests pass.}
    \label{fig:nulls}
\end{figure*}

We find the $\chi^2$ and null probabilities $p_0$ of the three frequency pairs (Table~\ref{tab:nulls}) to be
\begin{align}\label{eq:nulls_results}
    (95-150)\,\text{GHz:}&\quad \chi^2/N_{\rm dof}=1.41,\quad p_0=0.09, \nonumber\\
    (95-220)\,\text{GHz:}&\quad \chi^2/N_{\rm dof}=0.97,\quad p_0=0.51, \nonumber\\
    (150-220)\,\text{GHz:}&\quad \chi^2/N_{\rm dof}=0.79,\quad p_0=0.75.
\end{align}
All 24 bandpowers enter every difference, with the covariance evaluated at the velocity fit with a single $\bvel$. Every difference passes $p_0>0.05$ for both main and extended samples, and the largest excess is $1.3\sigma$, from the 95--150\,GHz pair. 
Figure~\ref{fig:nulls} shows the bandpower differences.

\subsubsection{Half-split null.}

In order to test that the map noise adds no dipole, we also measure the $\Pgv^{\ell=1}$ dipole of the extended sample, which contains every main-sample galaxy, on the half-difference of the two half-depth coadds of SPT-3G D1. We use the weights and filter of the full-depth maps. We estimate the covariance from 2000 surrogate realizations built on the half-difference maps, with no kSZ or foreground term and with the noise amplitudes $s_{nv}^\nu$ of the extended-sample velocity fit. The dipole is consistent with zero at 95, 150 and 220\,GHz ($\chi^2/N_{\rm dof}=1.21$, 0.81 and 0.93 for 24 bandpowers; $p_0=0.22$, 0.72 and 0.56).

\subsection{Bandpower correlations}\label{app:corr}

\begin{figure}[tbp]
    \centering
    \includegraphics[width=0.75\textwidth]{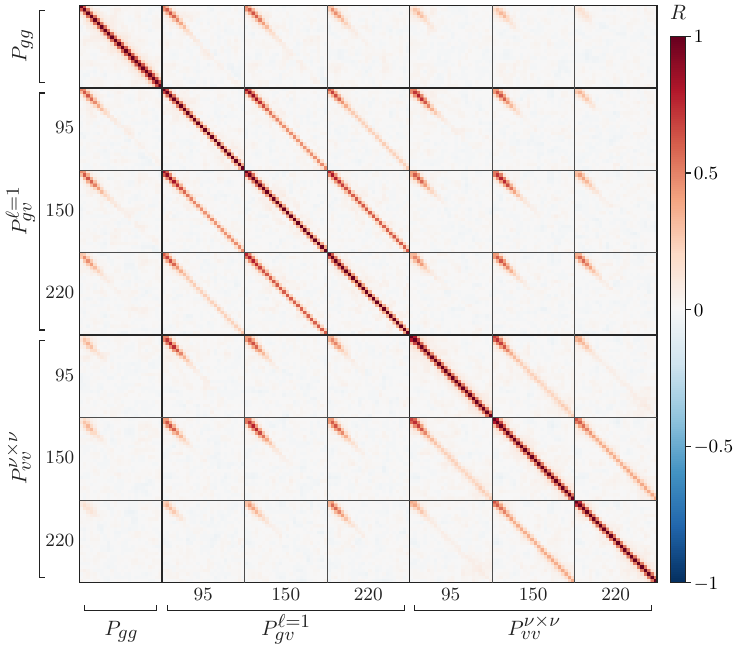}
    \vspace*{-12pt}
    \caption{Correlation coefficients $R_{ij}=\mathcal{C}_{ij}/(\mathcal{C}_{ii}\mathcal{C}_{jj})^{1/2}$ of the covariance $\mathcal{C}$ of the main-sample galaxy-galaxy, galaxy-velocity dipole and velocity-velocity surrogates at best-fit values at $\nu=95$, 150 and 220\,GHz.
    Within each block, $k$ increases to the right and downward over the 24 bandpowers, centred at $k=0.001$ to $0.047\,{\rm Mpc}^{-1}$.}
    \label{fig:corr}
\end{figure}

Finally, Figure~\ref{fig:corr} shows the correlation matrix of the covariance of the main-sample galaxy-velocity and velocity-velocity correlations.
Below $k=0.01\,{\rm Mpc}^{-1}$, the dipoles of two frequencies are correlated by up to 0.89 because they share the velocity signal.
Above $k=0.02\,{\rm Mpc}^{-1}$, the correlation between frequencies, in the dipoles and in the velocity auto-spectra, follows the correlation of the reconstruction noise, which the three maps share through the primary CMB and foreground anisotropies.

\section{Discussion}\label{sec:discussion}
We have measured for the first time the galaxy--velocity cross-power spectrum and the velocity auto-spectrum using SPT-3G D1 maps at 95, 150, and 220\,GHz and DESI-LS LRGs. Together with $P_{ge}$ from the same D1 maps and DESI-LS LRGs~\citep{Hotinli:2026zne}, both observables of the kSZ bispectrum~\citep{Smith:2018bpn} (the small-scale $P_{ge}$ and the large-scale $\Pgv$) have now been measured at high accuracy and in agreement with ACT results. 

For the main sample, we detect the kSZ signal in the galaxy-velocity dipole at SNR $10.4$ and in the velocity auto-spectrum at $7.2$, combining the three frequencies.
The velocity biases fitted at the three frequencies agree, and the frequency differences of the dipole are consistent with zero, as expected for the frequency-independent kSZ signal.
We measure the foreground bias from the galaxy-velocity monopole, which is dominated by the tSZ effect at 95 and 150\,GHz and by the CIB at 220\,GHz.

We constrain $\fnl$ to around $\pm60$ using the velocity-galaxy correlation and at around $\pm40$ combined with the galaxy auto-spectrum. Our $\fnl$ measurements are consistent with zero, as in previous ACT $\times$ DESI measurements~\citep{Hotinli:2025tul, Chaussidon:2026vmn}.
Across our alternative analysis choices, $\fnl$ moves by less than $0.4\sigma$ in both samples, and the extended sample agrees with the main sample.

The reconstruction recovers the radial velocity field up to the scale-independent factor $\bvel$, which we marginalize over in our $\fnl$ constraints.
The value of $\bvel$ depends on the fiducial galaxy-electron cross-spectrum $P_{ge}$ we assume on small scales, and is therefore also a probe of astrophysics.

Our halo-model--based calculation of $P_{ge}$ uses the Battaglia profile~\citep{Battaglia2016}, truncated at the virial radius and normalized so that the enclosed gas mass equals the cosmic baryon fraction $\Omega_b/\Omega_m$ of the halo mass; on large scales, the electrons then trace the matter. This construction largely fails to capture the reduction in $P_{ge}$ on small scales due to baryonic feedback seen in recent measurements~\citep[e.g.][]{Hotinli:2026zne, Qu:2026zyh, Hadzhiyska:2025mvt,Hadzhiyska:2024qsl,RiedGuachalla:2025byu}. As a result, we find $\bvel\sim0.25$; previous ACT$\times$DESI measurements~\citep{Hotinli:2025tul, Chaussidon:2026vmn} likewise found values below unity. The fit quality (SNRs, PTEs, and $p_0$ values) and our $\fnl$ constraints do not depend on this choice.

Furthermore, since $\bvel$ is roughly a weighted average over multipoles of the ratio $P_{ge}^{\rm true}/P_{ge}^{\rm fid}$, filters that weight scales differently can yield different values of $\bvel$. We apply a uniform filter across our SPT-3G D1 maps, but because SPT-3G has lower noise on small scales, this filter is centered at higher multipoles than the ACT filter used in~\citet{Hotinli:2025tul} (median $\ell\approx5100$ versus 3600). Baryonic feedback suppresses $P_{ge}^{\rm true}$ relative to $P_{ge}^{\rm fid}$ on small scales, so the extra weight our filter gives these scales lowers $\bvel$ further. We argue that this partly explains the difference between the $\bvel$ measurements in~\citet{Hotinli:2025tul} and in this work.

For the first time, we detect the velocity--galaxy cross-correlation monopole and dipole signals with the 220\,GHz CMB maps. We use the 220\,GHz monopole to separate and measure the tSZ and CIB contributions. We detect the CIB amplitude with monopole at over $2\sigma$ significance with the extended sample.   

Our frequency difference tests of the dipole are consistent with zero, as expected for the frequency-independent kSZ signal. Across alternative analysis choices we consider, we find $\fnl$ moves by less than $0.4\sigma$, and the extended sample agrees with the main sample. As in previous works~\citep{Hotinli:2025tul,Chaussidon:2026vmn}, we find that the galaxy--velocity cross-correlation is largely insensitive to imaging systematics.

Beyond $\fnl$~\citep{Munchmeyer2018}, reconstructed large-scale velocities can probe early-Universe physics including the primordial trispectrum amplitude $\tau_{\rm NL}$~\citep{AnilKumar:2022flx}, other primordial shapes~\citep{Adshead:2024paa}, compensated isocurvature perturbations~\citep{Hotinli:2019wdp,Kumar:2022bly}, pre-inflationary relics~\citep{Zhang:2015uta}, primordial vector fields~\citep{Vanzan:2023cze}, and CMB anomalies~\citep{Cayuso:2019hen}. Reconstructed velocities also provide late-time tests of dark energy~\citep{Adolff2025}, modified gravity~\citep{Pan:2019dax}, neutrino mass~\citep{Tishue2025} and relativistic effects~\citep{Contreras:2019bxy}, and probe the epoch of reionisation~\citep{Hotinli:2020csk}.
Velocity reconstruction has recently been used to bound an intrinsic dipole between the CMB and large-scale structure~\citep{Krywonos:2024mpb} and the cosmic bulk flow~\citep{Gandhi2026}.

Our results use the observations from 2019 and 2020, the first two seasons of the SPT-3G Main survey. The full survey depth will improve the velocity reconstruction, and the growing spectroscopic overlap with DESI will improve the photometric-redshift calibration. In the southern sky, the Rubin Observatory LSST~\citep{LSSTScience:2009jmu} and Euclid~\citep{Amendola:2016saw} will provide deeper galaxy samples for velocity reconstruction with SPT as the CMB counterpart; larger maps and catalogs over a wider area will strengthen the constraint on primordial non-Gaussianity~\citep{Munchmeyer2018} and beyond. A follow-up paper will combine these SPT-3G measurements of this paper with ACT$\times$DESI-LS measurements in the Galactic North.

\section*{Acknowledgments}

We thank Gil Holder for very useful discussions.
SCH was supported by the P.~J.~E.~Peebles Fellowship at Perimeter Institute.
SCH acknowledges support from grant NSF PHY-2309135 to the Kavli Institute for Theoretical Physics (KITP), where this work was initiated.
SCH also acknowledges the Aspen Center for Physics, where this work was further developed and which is supported by National Science Foundation grant PHY-2210452.
EC and SF are supported by Lawrence Berkeley National Laboratory and the Director, Office of Science, Office of High Energy Physics of the U.S. Department of Energy under Contract No.\ DE-AC02-05CH11231.
KMS was supported by an NSERC Discovery Grant, by the Daniel Family Foundation, and by the Centre for the Universe at Perimeter Institute. Research at Perimeter Institute is supported in part by the Government of Canada through the Department of Innovation, Science and Economic Development and by the Province of Ontario through the Ministry of Colleges and Universities.
The computations reported in this work were run on the high-performance computing cluster at Perimeter Institute.

The South Pole Telescope program is supported by the National Science Foundation (NSF) through awards OPP-1852617 and OPP-2332483. Partial support is also provided by the Kavli Institute of Cosmological Physics at the University of Chicago.

We acknowledge the use of public data from the SPT-3G D1 release and from the DESI Legacy Imaging Surveys, the latter made available through the National Energy Research Scientific Computing Center (NERSC), a U.S. Department of Energy Office of Science User Facility operated under Contract No.\ DE-AC02-05CH11231.

The Legacy Surveys consist of three individual and complementary projects: the Dark Energy Camera Legacy Survey (DECaLS; Proposal ID 2014B-0404; PIs: David Schlegel and Arjun Dey), the Beijing-Arizona Sky Survey (BASS; NOAO Prop. ID 2015A-0801; PIs: Zhou Xu and Xiaohui Fan), and the Mayall z-band Legacy Survey (MzLS; Prop. ID 2016A-0453; PI: Arjun Dey). DECaLS, BASS and MzLS together include data obtained, respectively, at the Blanco telescope, Cerro Tololo Inter-American Observatory, NSF's NOIRLab; the Bok telescope, Steward Observatory, University of Arizona; and the Mayall telescope, Kitt Peak National Observatory, NOIRLab. Pipeline processing and analyses of the data were supported by NOIRLab and the Lawrence Berkeley National Laboratory (LBNL). The Legacy Surveys project is honored to be permitted to conduct astronomical research on Iolkam Du'ag (Kitt Peak), a mountain with particular significance to the Tohono O'odham Nation. NOIRLab is operated by the Association of Universities for Research in Astronomy (AURA) under a cooperative agreement with the National Science Foundation. LBNL is managed by the Regents of the University of California under contract to the U.S. Department of Energy.

This project used data obtained with the Dark Energy Camera (DECam), which was constructed by the Dark Energy Survey (DES) collaboration. Funding for the DES Projects has been provided by the U.S. Department of Energy, the U.S. National Science Foundation, the Ministry of Science and Education of Spain, the Science and Technology Facilities Council of the United Kingdom, the Higher Education Funding Council for England, the National Center for Supercomputing Applications at the University of Illinois at Urbana-Champaign, the Kavli Institute of Cosmological Physics at the University of Chicago, Center for Cosmology and Astro-Particle Physics at the Ohio State University, the Mitchell Institute for Fundamental Physics and Astronomy at Texas A\&M University, Financiadora de Estudos e Projetos, Fundacao Carlos Chagas Filho de Amparo a Pesquisa do Estado do Rio de Janeiro, Conselho Nacional de Desenvolvimento Cientifico e Tecnologico and the Ministerio da Ciencia, Tecnologia e Inovacao, the Deutsche Forschungsgemeinschaft and the Collaborating Institutions in the Dark Energy Survey. The Collaborating Institutions are Argonne National Laboratory, the University of California at Santa Cruz, the University of Cambridge, Centro de Investigaciones Energeticas, Medioambientales y Tecnologicas-Madrid, the University of Chicago, University College London, the DES-Brazil Consortium, the University of Edinburgh, the Eidgenossische Technische Hochschule (ETH) Zurich, Fermi National Accelerator Laboratory, the University of Illinois at Urbana-Champaign, the Institut de Ciencies de l'Espai (IEEC/CSIC), the Institut de Fisica d'Altes Energies, Lawrence Berkeley National Laboratory, the Ludwig Maximilians Universitat Munchen and the associated Excellence Cluster Universe, the University of Michigan, NSF's NOIRLab, the University of Nottingham, the Ohio State University, the University of Pennsylvania, the University of Portsmouth, SLAC National Accelerator Laboratory, Stanford University, the University of Sussex, and Texas A\&M University.

BASS is a key project of the Telescope Access Program (TAP), which has been funded by the National Astronomical Observatories of China, the Chinese Academy of Sciences (the Strategic Priority Research Program ``The Emergence of Cosmological Structures'' Grant \# XDB09000000), and the Special Fund for Astronomy from the Ministry of Finance. The BASS is also supported by the External Cooperation Program of Chinese Academy of Sciences (Grant \# 114A11KYSB20160057), and Chinese National Natural Science Foundation (Grant \# 12120101003, \# 11433005). The Legacy Survey team makes use of data products from the Near-Earth Object Wide-field Infrared Survey Explorer (NEOWISE), which is a project of the Jet Propulsion Laboratory/California Institute of Technology. NEOWISE is funded by the National Aeronautics and Space Administration. The Legacy Surveys imaging of the DESI footprint is supported by the Director, Office of Science, Office of High Energy Physics of the U.S. Department of Energy under Contract No. DE-AC02-05CH11231, by the National Energy Research Scientific Computing Center, a DOE Office of Science User Facility under the same contract; and by the U.S. National Science Foundation, Division of Astronomical Sciences under Contract No. AST-0950945 to NOAO.

\section*{Data availability}

The SPT-3G D1 maps and ancillary products are publicly available through the SPT-3G data-release Globus collection ``SPT-3G Public Data Release'' (\texttt{0bf0cefa-568a-4b37-861d-32c9102ce7d6}, documented at \url{https://pole.uchicago.edu/public/data/quan26/index.html}; Ref.~\citep{SPT-3G:2026krg}), and the DESI Legacy Imaging Surveys DR9 LRG catalogues~\citep{Zhou:2023gji} and randoms through the DESI public data server (\url{https://data.desi.lbl.gov/public/papers/c3/lrg_xcorr_2023/} and \url{https://data.desi.lbl.gov/public/ets/target/catalogs/dr9/0.49.0/randoms/}). The modified version of the \texttt{kszx} package (\url{https://github.com/kmsmith137/kszx}) used in this work, together with the analysis code that produces the numbers and figures of this paper, will be made public upon publication and is available from the authors on request before then.

\appendix

\section{The extended sample}\label{app:extended}

In \S\ref{sec:results} we focused on the main galaxy sample of DESI-LS LRGs; here, we briefly go over our results for the extended sample that we skipped in the main part of this pair. Overall, we find that the main and extended samples agree very well with each other.

\begin{figure*}[tbp]
    \centering
    \includegraphics[width=\textwidth,trim={0 8 0 16},clip]{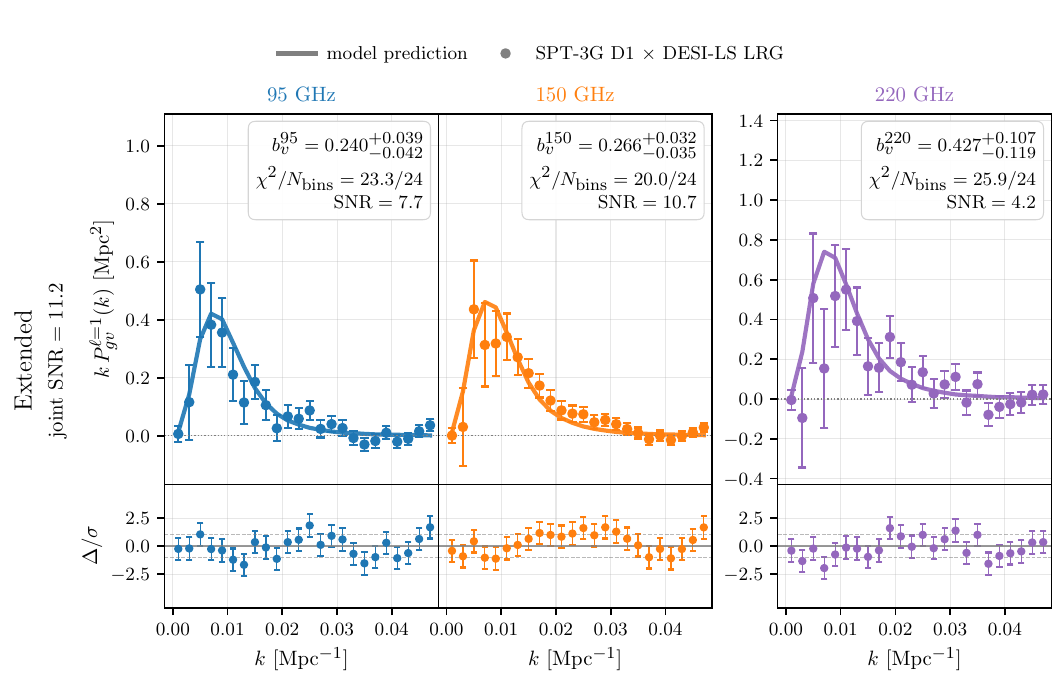}
    \vspace*{-24pt}
    \caption{As Fig.~\ref{fig:dipole_primary}, for the extended sample. The three dipoles give ${\rm PTE}_{\rm gof}=0.50$, 0.70 and 0.36 at 95, 150 and 220\,GHz.}
    \label{fig:dipole_ext}
\end{figure*}

We detect the galaxy-velocity correlation dipole for the extended sample at
\begin{equation}
\mathrm{SNR}~{\rm (velocity~dipole)}=
\begin{cases} 
    ~\,7.7\,~{\rm at~95\,GHz}\,,\\
    10.7\,~{\rm at~150\,GHz}\,,\\
    ~\,4.2\,~{\rm at~220\,GHz}\,,
    \end{cases}
\end{equation}
and we find the total SNR of $11.2$. Fitting the velocity bias values for each frequecy give 
\begin{equation}
b_v=
\begin{cases} 
    0.240^{+0.039}_{-0.042}\,~{\rm at~95\,GHz}\,,\\
    0.266^{+0.032}_{-0.035}\,~{\rm at~150\,GHz}\,,\\
    0.427^{+0.107}_{-0.119}\,~{\rm at~220\,GHz}.
    \end{cases}
\end{equation}
A single $\bvel$ shared accross frequencies gives the same SNR, and:
\begin{equation}\label{eq:bv_ext}
    \bvel=0.251^{+0.030}_{-0.032}\,,
\end{equation}
consistent with the per-frequency values. 
Figure~\ref{fig:dipole_ext} galaxy-velocity dipole bandpowers.

\begin{figure*}[tbp]
    \centering
    \includegraphics[width=\textwidth,trim={0 8 0 16},clip]{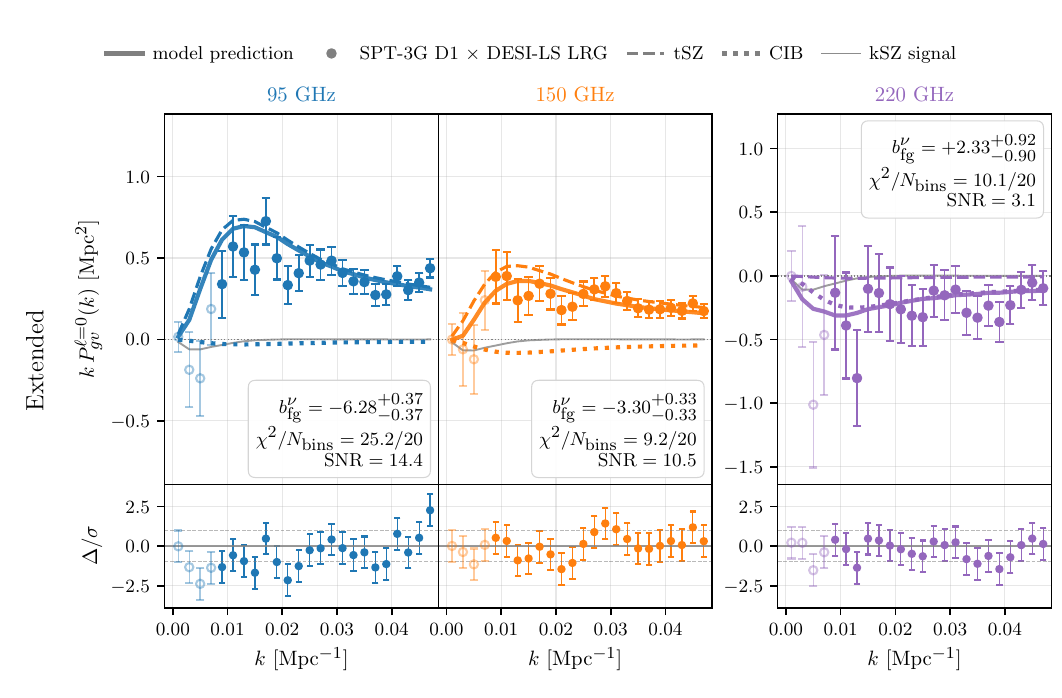}
    \vspace*{-24pt}
    \caption{As Fig.~\ref{fig:monopole_primary}, for the extended sample. The three monopoles give ${\rm PTE}_{\rm gof}=0.19$, 0.98 and 0.97 at 95, 150 and 220\,GHz.}
    \label{fig:monopole_ext}
\end{figure*}

For the foreground amplitudes, the galaxy-velocity monopole fit gives
\begin{equation}\label{eq:sed_amp_ext}
    A_{\rm tSZ} = (-4.01^{+0.23}_{-0.24})\times10^{3}\,\mu{\rm K}^{-1}, \qquad A_{\rm CIB} = (+0.72^{+0.30}_{-0.29})\times10^{3}\,\mu{\rm K}^{-1}\,
\end{equation}
 and the corresponding foreground biases of the extended sample are
\begin{equation}\label{eq:bfg_tot_ext}
    \bfg=\begin{cases}
    -6.3~{\rm at~95\,GHz}\,\\
    -3.3~{\rm at~150\,GHz}\,~\,\times10^3\mu{\rm K}^{-1}\,.\\
    +2.3~{\rm at~220\,GHz}\,\\        
    \end{cases}
\end{equation}
We detect foreground-induced monopole is detected at 
\begin{equation}
    {\rm SNR~(velocity~monopole)}=\begin{cases}
14.4~\rm at~95\,GHz\,,\\ 
10.5~\rm at~150\,GHz\,,\\
~\,3.1~\rm at~220\,GHz\,,
\end{cases}
\end{equation}
and the total SNR of $18.9$. Figure~\ref{fig:monopole_ext} shows the bandpowers with the model of the monopole fit.

\begin{figure*}[tbp]
    \centering
    \includegraphics[width=\textwidth,trim={0 8 0 16},clip]{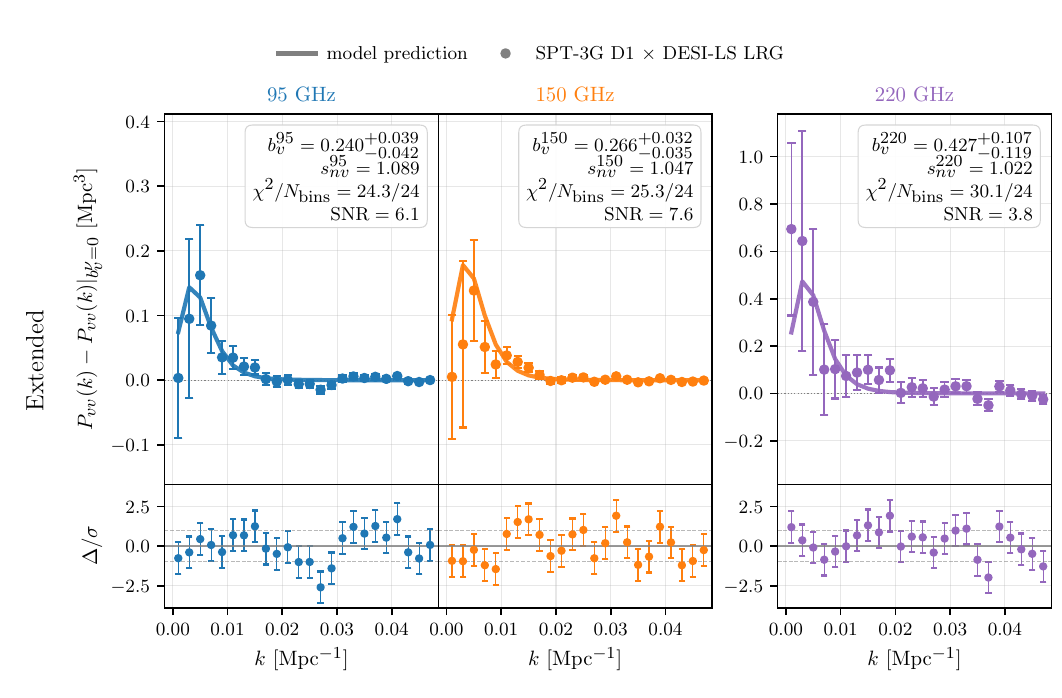}
    \vspace*{-24pt}
    \caption{As Fig.~\ref{fig:pvv_primary}, for the extended sample. Jointly, the auto-spectra reach kSZ SNR $8.1$. The three auto-spectra give ${\rm PTE}_{\rm gof}=0.44$, 0.39 and 0.18 at 95, 150 and 220\,GHz.}
    \label{fig:pvv_ext}
\end{figure*}

The reconstruction-noise amplitudes are 
\begin{equation}
    s_{nv}=\begin{cases}
        1.09~\rm at~95\,GHz\,,\\
        1.05~\rm at~150\,GHz\,,\\
        1.02~\rm at~220\,GHz\,,        
    \end{cases}
\end{equation}
each with uncertainty $0.01$. The per-frequency kSZ SNRs of the velocity auto-spectra are 
\begin{equation}
    {\rm SNR~(velocity~auto)}=\begin{cases}
        6.1~\rm at~95\,GHz\,,\\
        7.6~\rm at~95\,GHz\,,\\
        3.8~\rm at~95\,GHz\,,\\
    \end{cases}
\end{equation}
and their joint kSZ SNR is $8.1$. The auto-spectra give $\chi^2=\{24.3,25.3,30.1\}$ (24 bandpowers each) at the best fit, where ${\rm PTE}_{\rm gof}=\{0.44,0.39,0.18\}$. For the velocity damping scale, we find $\sigmav=5.5^{+5.7}_{-3.8}\,$Mpc. Figure~\ref{fig:pvv_ext} shows the noise- and foreground-subtracted bandpowers and the signal model.

For the extended sample, we find that the $P_{gv}^{\ell=1}+P_{vv}$ fits gives:
\begin{equation}\label{eq:joint_fnl_ext}
    \fnl=-58^{+69}_{-72}\,,
\end{equation}
where for a single $b_v$ shared between different frequencies, we find $\fnl=-50^{+63}_{-63}$, both consistent with zero. The full fit $(P_{gg}+P_{gv}^{\ell=1}+P_{vv})$ gives:
\begin{equation}\label{eq:full_fnl_ext}
    b_g=1.99^{+0.07}_{-0.06}\hspace{1.5cm}{\rm and}\hspace{1.5cm}\fnl=-40^{+41}_{-41}\,,
\end{equation}
and $\fnl=-38^{+40}_{-39}$ with a single $\bvel$. We give the $\fnl$ values from $P_{gg}$ alone in \S\ref{sec:results_pgg}.

\begin{figure*}[tbp]
    \centering
    \includegraphics[width=\textwidth,trim={0 4 0 11},clip]{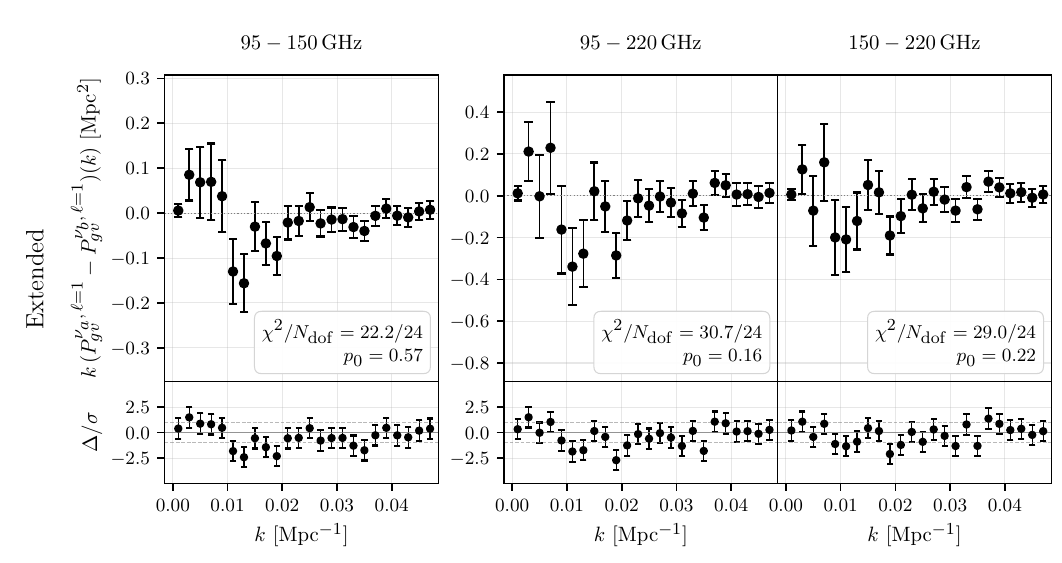}
    \vspace*{-24pt}
    \caption{As Fig.~\ref{fig:nulls}, for the extended sample. All three tests pass.}
    \label{fig:nulls_ext}
\end{figure*}

The three frequency differences of the extended sample are consistent with zero. The reduced $\chi^2$ and null probabilities $p_0$ (Table~\ref{tab:nulls}) are:
\begin{align}\label{eq:nulls_results_ext}
    (95-150)\,\text{GHz:}&\quad \chi^2/N_{\rm dof}=0.93,\quad p_0=0.57, \nonumber\\
    (95-220)\,\text{GHz:}&\quad \chi^2/N_{\rm dof}=1.28,\quad p_0=0.16, \nonumber\\
    (150-220)\,\text{GHz:}&\quad \chi^2/N_{\rm dof}=1.21,\quad p_0=0.22,
\end{align}
with the covariance evaluated at the velocity fit with a single $\bvel$. Concatenating 95--150 and 150--220\,GHz with their full cross-pair covariance gives $\chi^2/N_{\rm dof}=49.4/48$ ($p_0=0.42$). Figure~\ref{fig:nulls_ext} shows the difference null test bandpowers.

\begin{figure*}[tbp]
    \centering
    \includegraphics[width=0.78\textwidth,trim={0 12 0 0},clip]{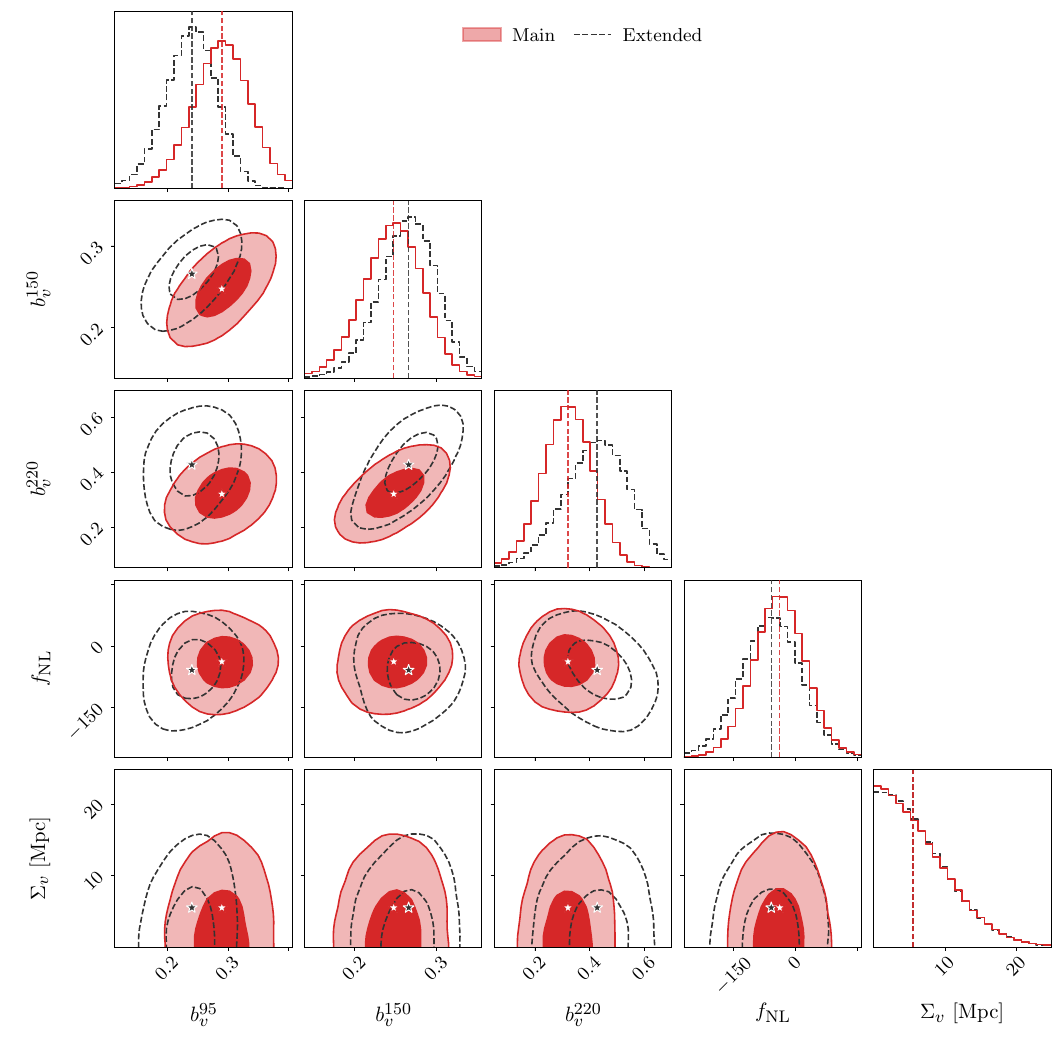}
    \vspace*{-12pt}
    \caption{Marginalized posteriors of the per-frequency velocity biases $\bvel^\nu$, $\fnl$ and the velocity damping scale $\sigmav$ from the velocity fit, for the main (red, filled) and extended (grey, dashed) samples. The reconstruction-noise amplitudes $s_{nv}^\nu$ are marginalized and not shown. The diagonal panels show the one-dimensional marginals; the contours enclose the $1\sigma$ and $2\sigma$ regions of a two-dimensional Gaussian ($39\%$ and $86\%$ of the two-dimensional posterior mass). Stars and dashed lines mark the posterior medians. The $\sigmav$ marginals peak at the lower prior edge, $\sigmav=0$.} 
    \label{fig:corner_primary}
\end{figure*}

Figure~\ref{fig:corner_primary} compares the posteriors of the main and extended samples. The $\fnl$ values of Eqs.~(\ref{eq:joint_fnl}) and~(\ref{eq:joint_fnl_ext}) differ by less than the error of either. Every main-sample galaxy belongs to the extended sample, so the errors are correlated, and we quote no significance for the difference. The tSZ amplitudes of the two samples agree within their errors, and we find $A_{\rm tSZ}^{\rm main}/A_{\rm tSZ}^{\rm ext} = 1.04$.

\bibliography{spt_paper_v2}

\end{document}

%% file: table_main.tex
\begin{table*}[t]
\centering
\caption{Parameter constraints from SPT-3G and DESI-LS LRG main and extended samples, as described in the main text.  
Rows correspond to our results from galaxy auto-spectrum, the galaxy-velocity monopole, galaxy-velocity dipole and velocity-velocity correlations. The last set of rows correspond to our joint fit of galaxy auto and galaxy-velocity dipole and velocity-velocity correlation.
Columns correspond to main and extended DESI-LS samples and two different parametrization of kSZ velocity bias: separate fits for each frequency, and a single $b_v$ for all frequency maps.
We give the details and figures of the extended sample in Appendix~\ref{app:extended}.}
\label{tab:main_consolidated}
\renewcommand{\arraystretch}{1.18}
\begin{tabular}{l l c c c c}
\hline\hline
 & & \multicolumn{2}{c}{Main} & \multicolumn{2}{c}{Extended} \\
Observable & Quantity & per-frequency $\bvel^\nu$ & single $\bvel$ & per-frequency $\bvel^\nu$ & single $\bvel$ \\
\hline
\multicolumn{6}{l}{\textit{Galaxy auto-spectrum} $\Pgg$: best fit; $b_g$, $s_n$ and $\Sigma_g$ held fixed in the monopole and velocity fits} 
\vspace*{3pt} \\
 & $b_g$ & \multicolumn{2}{c}{$2.186$} & \multicolumn{2}{c}{$2.009$} \\
 & $s_n$ (Poisson level) & \multicolumn{2}{c}{$1.025$} & \multicolumn{2}{c}{$1.082$} \\
 & $\Sigma_g$ [Mpc] & \multicolumn{2}{c}{$4.59$} & \multicolumn{2}{c}{$6.57$} \\
 & $\fnl$ (marginalised) & \multicolumn{2}{c}{$-33^{+42}_{-47}$} & \multicolumn{2}{c}{$-44^{+43}_{-45}$} \\
 & $\chi^2/N_{\rm dof}$ (PTE) & \multicolumn{2}{c}{$17.7/21\ (0.67)$} & \multicolumn{2}{c}{$12.3/21\ (0.93)$} \\
\hline
\multicolumn{6}{l}{\textit{Monopole fits} ($\Pgv^{\ell=0}$): foreground amplitudes [$10^{3}\,\mu{\rm K}^{-1}$] and goodness of fit} \vspace*{3pt} \\
 & $A_{\rm tSZ}$ & $-4.18^{+0.28}_{-0.28}$ & $-4.18^{+0.28}_{-0.28}$ & $-4.01^{+0.23}_{-0.24}$ & $-4.02^{+0.24}_{-0.24}$ \\
 & $A_{\rm CIB}$ & $+0.51^{+0.34}_{-0.34}$ & $+0.51^{+0.34}_{-0.33}$ & $+0.72^{+0.30}_{-0.29}$ & $+0.74^{+0.30}_{-0.30}$ \\
 & SNR, detection & $16.8$ & $16.8$ & $18.9$ & $18.8$ \\
 & SNR, amplitude & $15.5$ & $15.5$ & $17.0$ & $17.0$ \\
 & $\chi^2/N_{\rm dof}$ (PTE) & $38.6/58\ (0.98)$ & $38.6/58\ (0.98)$ & $51.6/58\ (0.71)$ & $51.4/58\ (0.72)$ \\
\hline
\multicolumn{6}{l}{\textit{Velocity fits} ($\Pgv^{\ell=1}$ dipole and $\Pvv$ auto-spectra,  $\bfg^\nu$ held fixed)} \vspace*{3pt} \\
 & $\bvel^{95}$ & $0.290^{+0.043}_{-0.046}$ & --- & $0.240^{+0.039}_{-0.042}$ & --- \\
 & $\bvel^{150}$ & $0.248^{+0.033}_{-0.036}$ & --- & $0.266^{+0.032}_{-0.035}$ & --- \\
 & $\bvel^{220}$ & $0.320^{+0.086}_{-0.087}$ & --- & $0.427^{+0.107}_{-0.119}$ & --- \\
 & single $\bvel$ & --- & $0.255^{+0.033}_{-0.035}$ & --- & $0.251^{+0.030}_{-0.032}$ \\
 & $\sigmav$ [Mpc] & $5.5^{+5.8}_{-3.9}$ & $5.0^{+5.6}_{-3.5}$ & $5.5^{+5.7}_{-3.8}$ & $5.4^{+5.8}_{-3.8}$ \\
 & $\fnl$ & $-38^{+61}_{-61}$ & $-35^{+62}_{-63}$ & $-58^{+69}_{-72}$ & $-50^{+63}_{-63}$ \\
 & $P_{gv}^{\ell=1}$ SNR, detection & $10.4$ & $10.4$ & $11.2$ & $11.3$ \\
 & $P_{vv}$ SNR, detection & $7.2$ & $6.9$ & $8.1$ & $7.9$ \\
 & $P_{gv}^{\ell=1}$ SNR, amplitude & $7.3$ & $7.4$ & $7.8$ & $7.7$ \\
 & $P_{vv}$ SNR, amplitude & $3.3$ & $3.3$ & $3.3$ & $3.2$ \\
 & $\chi^2/N_{\rm dof}$ (PTE) & $172.3/136\ (0.019)$ & $175.3/138\ (0.017)$ & $154.1/136\ (0.14)$ & $164.9/138\ (0.059)$ \\
\hline
\multicolumn{6}{l}{\textit{Full fit} ($\Pgg+\Pgv^{\ell=1}+\Pvv$)} \vspace*{3pt} \\
 & $\fnl$ & $-20^{+40}_{-46}$ & $-17^{+40}_{-46}$ & $-40^{+41}_{-41}$ & $-38^{+40}_{-39}$ \\
 & $b_g$ & $2.20^{+0.08}_{-0.06}$ & $2.20^{+0.08}_{-0.06}$ & $1.99^{+0.07}_{-0.06}$ & $2.00^{+0.07}_{-0.05}$ \\
 & $\chi^2/N_{\rm dof}$ (PTE) & $185.5/158\ (0.067)$ & $188.9/160\ (0.059)$ & $165.5/158\ (0.32)$ & $176.3/160\ (0.18)$ \\
\hline\hline
\end{tabular}
\end{table*}

%% file: table_nulls.tex
\begin{table*}[t]
\centering
\caption{SPT-3G $\times$ DESI-LS frequency-difference null tests on the $\Pgv^{\ell=1}$ dipole.}
\label{tab:nulls}
\renewcommand{\arraystretch}{1.16}
\begin{tabular}{llccc}
\toprule
Pair & Sample & $\chi^2/N_{\rm dof}$ & $p_0$ (single $b_v$)   \\
\midrule
\quad $95-150$ & Main & 33.8/24 & 0.08 (0.09)  \\
 & Extended  & 22.2/24 & 0.57 (0.57)  \\
\quad $95-220$ & Main  & 23.2/24 & 0.51 (0.51)  \\
 & Extended  & 30.7/24 & 0.16 (0.22)  \\
\quad $150-220$ & Main  & 19.1/24 & 0.75 (0.75)  \\
 & Extended  & 29.0/24 & 0.22 (0.29)  \\
\bottomrule
\end{tabular}
\end{table*}